\documentclass[12pt]{article}
\usepackage{graphicx,psfrag,epsf}
\usepackage{enumerate}
\usepackage{natbib}
\usepackage{url} % not crucial - just used below for the URL 
\usepackage{bm}
\usepackage{pbox}
\usepackage{verbatim}
\usepackage{fancyvrb}
\usepackage{multirow}
\usepackage{multicol}
\usepackage{ctable}
\usepackage{amssymb, amsthm, amsmath}
\usepackage{booktabs}
\usepackage{hyperref}
\usepackage{subfig}
\usepackage{rotating}
\usepackage{makecell}
\usepackage{xcolor}
\usepackage{cprotect}
\usepackage{array}
\usepackage[font=scriptsize,labelfont=bf]{caption}
\usepackage[export]{adjustbox}
\usepackage{changepage}
\usepackage{authblk}
\usepackage[top=1.8in,bottom = 0.5in,
left=1.6in,right=1.6in]{geometry}
\newcolumntype{T}[1]{>{\centering\let\newline\\\arraybackslash\hspace{0pt}}m{#1}}
\usepackage{lipsum}

\let\OLDthebibliography\thebibliography
\renewcommand\thebibliography[1]{
	\OLDthebibliography{#1}
	\setlength{\parskip}{0pt}
	\setlength{\itemsep}{0pt plus 0.3ex}
}

\newcommand{\RomanNumeralCaps}[1]
{\MakeUppercase{\romannumeral #1}}

\usepackage{xr}
\title{\bf Bayesian Modeling of Microbiome Data\\ for Differential Abundance Analysis}
\date{}
\author[1]{Qiwei Li \thanks{These authors contributed equally to this work. } }
\author[2, 4]{Shuang Jiang$^\ast$ }
\author[3]{Andrew Y. Koh }
\author[4]{Guanghua Xiao\thanks{To whom correspondence should be addressed.}}
\author[4]{Xiaowei Zhan $^{\dagger}$}
\affil[1]{Center for Depression Research and Clinical Care, Department of Psychiatry, University of Texas Southwestern Medical Center}
\affil[2]{Department of Statistical Science, Southern Methodist University}
\affil[3]{Departments of Pediatrics and Microbiology, University of Texas Southwestern Medical Center}
\affil[4]{Quantitative Biomedical Research Center, Department of Population and Data Sciences, University of Texas Southwestern Medical Center}
\graphicspath{{figs/}}

\begin{document}
	\maketitle
	
	\setlength{\abovedisplayskip}{3pt}
	\setlength{\belowdisplayskip}{3pt}
	
	\DefineShortVerb{\#}% # denotes verbatim opening/closing character
	\SaveVerb{deseq2}#DeSeq2#
	\SaveVerb{edger}#edgeR#
	\SaveVerb{metagenomeSeq}#metagenomeSeq#

	\bigskip
	\begin{abstract}
	Advances in next-generation sequencing technology have enabled the high-throughput profiling of metagenomes and accelerated the study of the microbiome. Recently, there is a rise of numerous studies that aim to decipher the relationship between the microbiome and disease. One of the most essential questions is to identify differentially abundant taxonomic features across different populations (such as cases and controls). Microbiome count data are high-dimensional and usually suffer from uneven sampling depth, over-dispersion, and zero-inflation. These characteristics often hamper downstream analysis and thus require specialized analytical models. Here we propose a general Bayesian framework to model microbiome count data for differential abundance analysis. This framework is composed of two hierarchical levels. The bottom level is a multivariate count-generating process from multiple choices. We particularly focus on the choice of a zero-inflated negative binomial (ZINB) model that takes into account the skewness and excess zeros in the microbiome data and incorporates model-based normalization through prior distributions with stochastic constraints. The top level is a mixture of Gaussian distributions with a feature selection scheme, which enables us to identify a set of differentially abundant taxa. In addition, the model allows us to incorporate phylogenetic tree information into the framework via the use of Markov random field priors. All the parameters are simultaneously inferred by using Markov chain Monte Carlo sampling techniques. Comprehensive simulation studies are conducted to evaluate our method and compared it with alternative approaches. Applications of the proposed method to two real microbiome datasets show that our method is able to detect a set of differentially abundant taxa at different taxonomic ranks, most of which have been experimentally verified. In summary, this statistical methodology provides a new tool for facilitating advanced microbiome studies and elucidating disease etiology.
\end{abstract}
	
	\noindent%
	\vfill
	
	\newpage

\section{Introduction}
The human body hosts more than $100$ trillion microorganisms. Collectively, the microorganism genomes contains at least $100$ times as many genes as the human genome \citep{backhed2005host}. The microbes in a healthy body aid in digestion and metabolism, and prevent the colonization of pathogenic microorganisms \citep{honda2012microbiome}. However, an impaired microbiome has been found to be associated with a number of human diseases, such as liver cirrhosis \citep{zeller2014potential}, schizophrenia \citep{castro2015composition}, etc. Accurate identification of microbiota-disease associations could facilitate the elucidation of disease etiology and lead to novel therapeutic approaches.

Advances in next-generation sequencing (NGS) technology, such as high-throughput 16S rRNA gene and metagenomic profiling, have accelerated microbiome research by generating enormous amounts of low-cost sequencing data \citep{metzker2010sequencing}. The availability of massive data motivates the development of specialized analytical models to identify disease-associated microbiota, for example, a set of taxa whose abundances significantly differ across clinical outcomes. Perhaps the simplest approach is to first convert the count data to its compositional version via dividing each read count by the total number of reads in each sample, and then apply the Wilcoxon rank-sum test (or its generalized version,  the Kruskal–Wallis test) to each taxonomic feature individually \citep{la2015hypothesis}. Differential abundance analysis has also been extensively studied in other types of NGS data. Hence, several methods that were designed for RNA-Seq, including \texttt{edgeR} \citep{robinson2010edger}, \texttt{DESeq2} \citep{love2014moderated}, and their modifications \citep{mandal2015analysis}, have been used for analyzing microbiome count data. 
However, those methods result in strong biases since they neglect to account for the excess zeros observed in the microbiome data. The sparsity is usually due to both biological and technical phenomena: some microorganisms are found in only a small percentage of samples, whereas others are simply not detected owing to insufficient segueing depth \citep{Paulson2013}. 

Recently, a number of zero-inflated (ZI) models have been proposed to analyze zero-inflated microbiome count data, including the ZI Gaussian model \citep{Paulson2013}, ZI negative binomial model \citep{zhang2016zero}, and ZI beta regression model \citep{peng2016zero}. \cite{xu2015assessment} argued that these models have an advantage in controlling type \RomanNumeralCaps{1} error. However, all of them require an \textit{ad hoc} normalizing factor for each sample to reduce biases due to uneven sequencing depth. From a statistical perspective, the employment of pre-normalized quantities leads to non-optimal performance and limits the power of downstream analysis \citep{mcmurdie2014waste}. In addition, the microbiome count data are also highly variable, both with respect to the number of total reads per sample and per taxonomic features. Hence, the distributions of observed counts are typically skewed and over-dispersed, since a large number of taxa are recorded at low frequencies whereas a few are recorded very frequently. In order to take account of this characteristic, statistical models based on either negative binomial or Dirichlet-multinomial distribution have been developed \citep{holmes2012dirichlet, Paulson2013,  chen2013variable,zhang2017negative}.

To overcome the aforementioned limitations: 1) zero-inflation, 2) uneven sampling depth, and 3) over-dispersion, we present a general Bayesian hierarchical framework to model microbiome count data for differential abundance analysis. It consists of two levels in order to allow flexibility. The bottom level is a multivariate count variable generating process, where a wide range of classic models can be plugged in, such as Dirichlet-multinomial (DM) model and zero-inflated negative binomial (ZINB) model. The mean parameters of the bottom-level model typically refer to the latent relative abundance. For the ZINB model, we further incorporate model-based normalization through Bayesian nonparametric prior distributions with stochastic constraints to infer the normalizing factors (i.e. sequencing depth). The top level is a mixture of Gaussian distributions to model the latent relative abundance with a feature selection scheme, which enables to identify a set of discriminatory taxa among different clinical groups. In addition, we introduce how to incorporate the phylogenetic structure to jointly select biologically similar taxa. In comprehensive simulation studies using both simulated and synthetic data, the proposed method outperforms the alternative approaches. The applications to two real microbiome datasets from a cancer study and a psychiatry study further demonstrated the advantage of the proposed method.

The rest of the paper is organized as follows. Section \ref{model} introduces the bi-level Bayesian modeling framework and discusses the prior formulations. Section \ref{posterior} briefly describes the Markov chain Monte Carlo algorithm and the resulting posterior inference. In Section \ref{simulation}, we present comprehensive simulation studies using both simulated and synthetic data to illustrate the performance of the method. Section \ref{realdata} consists of two case studies, using the proposed ZINB model. Section \ref{conclusion} concludes the paper with remarks on future directions.

\section{Model}\label{model}
We present a bi-level Bayesian framework for microbial differential abundance analysis. Section \ref{Count_generating_process} introduces two representative count generative models as the first (or bottom) level, while Section \ref{Gaussian_mixture} describes a Gaussian mixture model as the second (or top) level. Figure \ref{f1} in the supplement shows the graphical formulations of the proposed models. Before introducing the main components, we depict the input of our framework as follows.

Let $\bm{Y}$ denote an $n$-by-$p$ taxonomic abundance table of $n$ subjects and $p$ taxa, with $y_{ij}\in\mathbb{N},i=1,\ldots,n,j=1,\ldots,p$ indicating the count of taxon $j$ observed from subject $i$. Note that $\bm{Y}$ can be obtained from either 16S rRNA gene sequencing or the metagenomic shotgun sequencing (MSS). For the sake of simplicity, we assume that the taxonomic features in $\bm{Y}$ are all at the lowest available hierarchical levels (i.e. genus or operational taxonomic unit (OTU) for 16S rRNA data, and species for MSS data). As the count matrix at a higher taxonomic level can be easily summed up from its lower level, we discuss how to integrate information from a phylogenetic tree in Section \ref{tree}. We use an $n$-dimensional vector $\bm{z}=(z_1,\ldots,z_n)^T$ to allocate the $n$ subjects into $K$ different groups (i.e. phenotypes, conditions, etc.), with $z_i=k,k=1,\ldots,K$ indicating that subject $i$ belongs to group $k$. In addition, we use the following notations throughout this paper. For any $n$-by-$p$ matrix $\bm{X}$, we use $\bm{x}_{i\cdot}=(x_{i1},\ldots,x_{ip})^T$ and $\bm{x}_{\cdot j}=(x_{1j},\ldots,x_{nj})^T$ to denote  the vector from $i$-th row and $j$-th column of $\bm{X}$, and use $X_{i\cdot}=\sum_{j=1}^px_{ij}$ and $X_{\cdot j}=\sum_{i=1}^nx_{ij}$ to denote the sum of all counts in the $i$-th row and $j$-th column of $\bm{X}$.

\subsection{Multivariate count variable generating processes}\label{Count_generating_process}
In the bottom-level of the framework, we consider the multivariate counts in subject $i$, i.e. $\bm{y}_{i\cdot}$, as sampled from a probabilistic model $\mathcal{M}$. The model learns the latent relative abundance of each taxon in each subject, and should characterize one or more attributes of microbiome count data. More importantly, it automatically accounts for measurement errors and uncertainties associated with the counts \citep{li2015microbiome}. Without loss of generality, we write
\begin{equation}\label{top}
\bm{y}_{i\cdot}\sim\mathcal{M}(\bm{\alpha}_{i\cdot},\bm{\Theta}),
\end{equation}
where the positive vector $\bm{\alpha}_{i\cdot}=(\alpha_{i1},\ldots,\alpha_{ip})^T,\alpha_{ij}>0$ denotes the latent relative abundance for each taxon in subject $i$, and $\bm{\Theta}$ denotes all other model parameters. Table \ref{Table 1} provides a list of $\mathcal{M}$ and their features, two of which are discussed in detail as below.

\subsubsection{Dirichlet-multinomial model}\label{DM}
One commonly used candidate of $\mathcal{M}$ is the Dirichlet-multinomial (DM) model \citep[see e.g.][]{La2012,holmes2012dirichlet,chen2013variable,wadsworth2017integrative}. To illustrate the model, we start by modeling the counts observed in subject $i$ with a multinomial distribution $\bm{y}_{i\cdot}|\bm{\psi}_{i\cdot}\sim\text{Multi}(Y_{i\cdot},\bm{\psi}_{i\cdot})$. The $p$-dimensional vector $\bm{\psi}_{i\cdot}=(\psi_{i1},\ldots,\psi_{ip})^T$ is defined on a $p$-dimensional simplex (i.e. $\psi_{ij}>0,\forall j$ and $\sum_{j=1}^p\psi_{ij}=1$), and represents the underlying taxonomic abundances. The p.m.f. is $Y_{i\cdot}\prod_{j=1}^p\psi_{ij}^{y_{ij}}/y_{ij}!$, with the mean and variance of each component, $\text{E}(Y_{ij})=\psi_{ij}Y_{i\cdot}$ and $\text{Var}(Y_{ij})=\psi_{ij}(1-\psi_{ij})Y_{i\cdot}$, respectively. 

We further impose a Dirichlet prior on the multinomial parameter vector to allow for over-dispersed distributions, $\bm{\psi}_{i\cdot}|\bm{\alpha}_{i\cdot}\sim\text{Dir}(\bm{\alpha}_{i\cdot})$, where each element of the $p$-dimensional vector $\bm{\alpha}_{i\cdot}=(\alpha_{i1},\ldots,\alpha_{ip})^T$ is strictly positive. Due to the conjugacy between the Dirichlet distribution and the multinomial distribution, we can integrate $\bm{\psi}_{i\cdot}$ out, $p(\bm{y}_{i\cdot}|\bm{\alpha}_{i\cdot})=\int p(\bm{y}_{i\cdot}|\bm{\psi}_{i\cdot})p(\bm{\psi}_{i\cdot}|\bm{\alpha}_{i\cdot})d\bm{\psi}_{i\cdot}$, resulting in a DM model:  $\bm{y}_{i\cdot}|\bm{\alpha}_{i\cdot}\sim\text{DM}(\bm{\alpha}_{i\cdot}),
$ with the p.m.f. $f_\text{DM}(\bm{y}_{i\cdot}|\bm{\alpha}_{i\cdot})=\frac{\Gamma(Y_{i\cdot}+1)\Gamma(A_{i\cdot})}{\Gamma(Y_{i\cdot}+A_{i\cdot})}\prod_{j=1}^p\frac{\Gamma(y_{ij}+\alpha_{ij})}{\Gamma(y_{ij}+1)\Gamma(\alpha_{ij})}$, where $Y_{i\cdot}=\sum_{j=1}^{p}y_{ij}$ and $A_{i\cdot}=\sum_{j=1}^{p}a_{ij}$. The variance of each count variable is $\text{Var}(Y_{ij})=(Y_{i\cdot}+A_{i\cdot})/(1+A_{i\cdot})\text{E}(\psi_{ij})(1-\text{E}(\psi_{ij}))Y_{i\cdot}$. Comparing this with the multinomial model, we see that the variance of the DM is inflated by a factor of $(Y_{i\cdot}+A_{i\cdot})/(1+A_{i\cdot})$, Thus, the DM distribution can explicitly model extra variation. Note that $A_{i\cdot}=\sum_{j=1}^p\alpha_{ij}$ controls the degree of over-dispersion. A small value of $A_{i\cdot}$ results in large over-dispersion, while a large value approaching infinity reduces the DM model to a multinomial model.

\subsubsection{Zero-inflated negative binomial model}\label{ZINB}
Although the DM model offers more flexibility than the multinomial model in terms of modeling over-dispersion, neither models accounts for zero-inflation. The excess zeros are often attributed to rare or low abundance microbiota species that may be present in only a small percentage of samples, whereas others are not recorded owing to the limitations of the sampling effort. Thus, we consider modeling each taxonomic count using a zero-inflated negative binomial (ZINB) model,
\begin{equation}\label{zinb}
y_{ij}\sim \pi_i\text{I}(y_{ij} = 0)+(1-\pi_i)\text{NB}(\lambda_{ij},\phi_{j}),
\end{equation}
where we constrain one of the two mixture kernels to be degenerate at zero, thereby allowing for zero-inflation. In model (\ref{zinb}), $\pi_i\in(0,1)$ can be viewed as the proportion of extra zero counts in sample $i$. Here we use $\text{NB}(\lambda,\phi),\lambda,\phi>0$ to denote a negative binomial (NB) distribution, with expectation $\lambda$ and dispersion $1/\phi$. With this parameterization of the NB model, the p.m.f. is written as $\frac{\Gamma(y+\phi)}{y!\Gamma(\phi)}\left(\frac{\phi}{\lambda+\phi}\right)^\phi\left(\frac{\lambda}{\lambda+\phi}\right)^y$, with the variance $\text{Var}(Y)=\lambda+\lambda^2/\phi$. Note that $\phi$ controls the degree of over-dispersion. A small value indicates a large variance to mean ratio, while a large value approaching infinity reduces the NB model to a Poisson model with the same mean and variance. Now we rewrite model (\ref{zinb}) by introducing a latent indicator variable $\eta_{ij}$, which follows a Bernoulli distribution with parameter $\pi_i$, such that if $\eta_{ij}=1$ then $y_{ij}=0$, whereas if $\eta_{ij}=0$ then $y_{ij}\sim\text{NB}(\lambda_{ij},\phi_{j})$. The independent Bernoulli prior assumption can be further relaxed by formulating a $\text{Be}(a_\pi,b_\pi)$ hyperprior on $\pi_i$, leading to a beta-Bernoulli prior of $\eta_{ij}$ with expectation $a_\pi/(a_\pi+b_\pi)$. Setting $a_\pi=b_\pi=1$ results in a non-informative prior on $\pi_i$. Lastly, we specify the same prior distribution for each dispersion parameter as $\phi_j\sim\text{Ga}(a_\phi,b_\phi)$. Small values, such as $a_\phi=b_\phi=0.001$, result in a weakly informative gamma prior.

Multiplicative characterizations of the NB (or Poisson as a special case) mean are typical in both the frequentist \citep[e.g.][]{Witten2011,Li2012,Cameron2013} and the Bayesian literature \citep[e.g.][]{Banerjee2014,Airoldi2016} to justify latent heterogeneity and over-dispersion in multivariate count data. Here, we parameterize the mean of the NB distribution as the multiplicative effect of two parameters, $\lambda_{ij}=s_i\alpha_{ij}$. We denote $s_i$ as the size factor of sample $i$, reflecting the fact that samples are sequenced in different depths. Once this global effect is accounted for, $\alpha_{ij}$ is interpreted as the normalized abundance for counts $y_{ij}$. Conditional on the parameters, the likelihood of observing the counts $\bm{y}_{i\cdot}$ can be written as
\begin{equation}\label{zinb_likelihood} f_\text{ZINB}(\bm{y}_{i\cdot}|\bm{\alpha}_{i\cdot},\bm{\eta}_{i\cdot},\bm{\phi},s_i)=\prod_{j=1}^p\text{I}(y_{ij}=0)^{\eta_{ij}}\left(\frac{\Gamma(y_{ij}+\phi_j)}{y_{ij}!\Gamma(\phi_j)}\left(\frac{\phi_j}{s_i\alpha_{ij}+\phi_j}\right)^{\phi_j}\left(\frac{s_i\alpha_{ij}}{s_i\alpha_{ij}+\phi_j}\right)^{y_{ij}}\right)^{1-\eta_{ij}}.
\end{equation}

To ensure identifiability between the latent relative abundance $\alpha_{ij}$ and its relevant size factor $s_i$, one typical choice is to calculate $\bm{s}=(s_1,\ldots,s_n)$ based on the observed counts $\bm{Y}$, combined with some constraint such as $\sum_{i=1}^ns_i=1$ or $\prod_{i=1}^ns_i=1$ (i.e. $\sum_{i=1}^n\log s_i=0$). Table \ref{Table 2} summarizes the existing methods for estimating the size factors. The simplest approach is to set the size factor $s_i$ proportional to the total sum of counts in the sample, i.e. $\hat{s_i}\propto Y_{i\cdot}$, although it does not account for heteroscedasticity and yields biased estimation on all other model parameters \citep{dillies2013comprehensive}. In practice, most methods have been developed in the context of RNA-Seq data analyses. For example, in order to mitigate the influence of extremely low and high counts on the size factor estimation, \cite{Bullard2010} suggest matching the between-sample distributions in terms of their upper-quartiles (Q75) to normalize the counts from mRNA-Seq experiments. Furthermore, \cite{Anders2010} and \cite{Robinson2010} propose normalization techniques based on relative log expression (RLE) and weighted trimmed mean by M-values (TMM), respectively. Both of them assume that most features (e.g., genes) are not differentially abundant, and of those that are, there is an approximately balanced amount of increased/decreased abundance. However, these assumptions are likely not appropriate for highly diverse microbial environments \citep{weiss2017normalization}. %\cite{Witten2011} compared the clustering performance under different choices of the above size factor inferences with a constraint of $\sum_{i=1}^ns_i=1$. 
\cite{Paulson2013} developed a so-called cumulative sum scaling (CSS) method. It is an adaptive extension of Q75, which is better suited for microbiome data. While convenient, the use of the plug-in estimates $\hat{s}_i$ has noticeable shortcomings. In a Bayesian framework, those plug-in estimates can be viewed as point mass priors. On one hand, the ``double dipping" problem occurs as those informative priors are derived from the data before model fitting and thus the uncertainty quantification for estimation of $s_i$ will not be reflected in the inference; on the other hand, a discontinuity on the point mass priors may introduce bias in model parameter inference. To address the identifiability issue and allow flexibility in the estimation of the unknown normalizing factors $s_i$, \cite{Li2017} imposed a regularizing prior with a stochastic constraint on the logarithmic scale of each size factor. They assumed that $\log s_i$ is drawn from a mixture of a two-component Gaussian mixture,
\begin{equation}
\begin{split}
\log{s}_i \sim \sum_{m=1}^M\psi_m \left[t_m\, \text{N}(\nu_m,\sigma_s^2)+(1-t_m)\, \text{N}\left(-\frac{t_m\nu_m}{1-t_m},\sigma_s^2\right)\right],
\label{Equation221}
\end{split}
\end{equation}
with the weight of outer mixtures denoted by $\psi_m$ ($0<\psi_m<1$, $\sum_{m=1}^M\psi_m=1$), where $M$ is an arbitrary large positive integer. The use of mixture distributions allows for flexible estimation of the posterior density of $\log{s}_i$. In order to satisfy the desired stochastic constraint, each of $M$ components is further modeled by a mixture of two Gaussian distributions with a constant mean of zero. The weight of each inner mixture is denoted by $t_m$ ($0<t_m<1$). Note that if $M\rightarrow\infty$, model (\ref{Equation221}) can be interpreted as Bayesian nonparametric infinite mixtures. With the assumption that the weights $\psi_m$ are defined by the stick-breaking construction, i.e. $\psi_1 = V_1, \psi_m = V_m\prod_{u = 1}^{m-1}(1 - V_u)$, $m = 1, 2, \ldots$, it becomes a case of Dirichlet process mixture models, which have been extensively used in recent literature for flexible density estimation \citep[see][]{trippa2011, Kyung2011, taddy2012}. \cite{Lee2018} have demonstrated the superiority of employing the Dirichlet process prior (DPP) in a Bayesian semiparametric regression model for joint analysis of ocean microbiome data. It is said that DPP can accommodate various features in a distribution, such as skewness or multi-modality, while satisfying the mean constraint. We conclude the ZINB model by specifying the following hyper-prior distributions for DPP: $\nu_m \sim \text{N}(0, \tau_{\nu})$, $t_m \sim \text{Be}(a_t,b_t)$, and $V_m \sim \text{Be}(a_m, b_m)$. Note that $\psi_m$ will be updated according to the stick-breaking construction. We assume that $\sigma_s^2 = 1$, which completes an automatic normalization of the size factors.

%\vspace*{-0.8cm}
\subsection{Gaussian mixture models with feature selection}\label{Gaussian_mixture}
In the top level of our framework, we aim to identify a subset of taxa that are relevant to discriminating the $n$ subjects into $K$ distinct groups. We postulate the existence of a latent binary vector $\bm{\gamma}=(\gamma_1,\ldots,\gamma_p)^T$, with $\gamma_j=1$ if taxon $j$ is differentially abundant among the $K$ groups, and $\gamma_j=0$ otherwise. This assumption could be formulated as,
\begin{equation}\label{zinb_level2}
\log \alpha_{ij}|\gamma_j\sim
\begin{cases}
{\begin{array}{ll}
	\text{N}(\mu_{kj},\sigma_{kj}^2) & \text{ if }  \gamma_j=1 \text{ and } z_i=k\\
	\text{N}(\mu_{0j},\sigma_{0j}^2) & \text{ if } \gamma_j=0\\
	\end{array}}.
\end{cases}
\end{equation}
Note that the use of $\log$ transformation has two folds: 1) it ensures that the latent relative abundance $\alpha_{ij}$'s are not skewed; 2) it converts a positive value of $\alpha_{ij}$ to be either positive or negative, which is more appropriate for Gaussian fittings. A common choice for the prior of the binary latent vector $\bm{\gamma}$ is independent Bernoulli distributions on each individual component with a common hyperparameter $\omega$, i.e. $\gamma_j\sim\text{Bernoulli}(\omega)$. It is equivalent to a binomial prior on the number of discriminatory taxa, i.e. $p_\gamma=\sum_{j=1}^p\gamma_j\sim\text{Bin}(p,\omega)$. The hyperparameter $\omega$ can be elicited as the proportion of taxa expected \textit{a priori} to be differentially abundant among the $K$ groups. This prior assumption can be further relaxed by formulating a $\text{Be}(a_\omega,b_\omega)$ hyperprior on $\omega$, which leads to a beta-binomial prior on $p_\gamma$ with expectation $pa_\omega/(a_\omega+b_\omega)$. \cite{Tadesse2005} suggest a vague prior of $\omega$, by imposing the constraint $a_\omega+b_\omega=2$. %As a result, the desirable mean percentage of inclusion is $a_\omega/2$.

Taking a conjugate Bayesian approach, we impose a normal prior on $\mu_0$ and each $\mu_k$, and an inverse-gamma (IG) prior on $\sigma_0^2$ and each $\sigma_k^2$; that is, $\mu_{0j} \sim\text{N}(0,h_0\sigma_0^2)$, $\mu_{kj} \sim\text{N}(0,h_k\sigma_k^2)$, $\sigma_{0j}^2 \sim\text{IG}(a_0,b_0)$, and $\sigma_{kj}^2 \sim\text{IG}(a_k,b_k)$. This parametrization setting is standard in most Bayesian normal models. It allows for creating a computationally efficient feature selection algorithm by integrating out means (i.e. $\mu_0$ and $\mu_k$) and variances (i.e. $\sigma_0^2$ and $\sigma_k^2$). The integration leads to marginal non-standardized Student's t-distributions on $\log\alpha_{ij}$. Consequently, we can write the likelihood of observing the latent relevant abundances of taxon $j$ as,
\begin{equation}
\begin{split}
&p(\bm{\alpha}_{\cdot j}|\gamma_j)=(2\pi)^{-\frac{n}{2}}\times\\
&\begin{cases}
{\begin{array}{ll}
	\prod_{k=1}^K(n_kh_k+1)^{-\frac{1}{2}}\frac{\Gamma\left(a_k+\frac{n_k}{2}\right)}{\Gamma(a_k)}\frac{b_k^{a_k}}{\left\{b_k+\frac{1}{2}\left[\sum_{\{i:z_i=k\}}\log\alpha_{ij}^2-\frac{\left(\sum_{\{i:z_i=k\}}\log\alpha_{ij}\right)^2}{n_k+\frac{1}{h_k}}\right]\right\}^{a_k+\frac{n_k}{2}}} & \text{ if }  \gamma_j=1\\
	(nh_0+1)^{-\frac{1}{2}}\frac{\Gamma\left(a_0+\frac{n}{2}\right)}{\Gamma(a_0)}\frac{b_0^{a_0}}{\left\{b_0+\frac{1}{2}\left[\sum_{i=1}^n\log\alpha_{ij}^2-\frac{\left(\sum_{i=1}^n\log\alpha_{ij}\right)^2}{n+\frac{1}{h_0}}\right]\right\}^{a_0+\frac{n}{2}}} & \text{ if } \gamma_j=0\\
	\end{array}},
\end{cases}
\end{split}
\end{equation} 
where $n_k$ is the number of subjects belonging to group $k$. To specify the IG hyperparameters of $\sigma_0^2$ and $\sigma_k^2$, we recommend a weakly informative choice by setting the shape parameters $a_0$ and $a_k$'s to $2$, and the scale parameters $b_0$ and $b_k$ to $1$, following \cite{li2018bayesian}. To specify the hyperprior on $h_0$, we suggest setting it to a large value so as to obtain a fairly flat distribution over the region where the data are defined. According to \cite{stingo2013integrative}, a large value of $h_k$ allows for mixtures with widely different component means and typically encourages the selection of relatively large effects (e.g., those taxa of large effect size among groups), whereas a small value encourages the selection of small effects. We carried out a sensitivity analysis on the simulated data and found that  both DM and ZINB models perform reasonably well if the value ranges from $10$ to $100$.

\subsection{Incorporating the phylogenetic tree}\label{tree}
One feature of microbiome data is that the count matrix can be summarized at different taxonomic levels, since there is a natural hierarchy of biological organism classification, i.e. species, genus, family, order, class, etc. Given a count table $\bm{Y}$ at the most bottom-most level, we can aggregate the counts into any upper level based on the phylogenetic tree. A tree is an undirected graph where any two vertices are connected by exactly one path. Thus, we describe the phylogenetic tree by using the adjacent matrix in graph theory. Suppose the relationship between taxa in different levels are represented by a $p'\times p'$ symmetric matrix $\bm{G}$, with $g_{jj'}=1$ if taxon $j$ and $j'$ have a direct link in the tree. Let $l,1\le l\le L$ index the taxonomic level in the order of $\{\text{species},\text{genus},\text{family},\text{order},\text{class},\text{phylum},\text{kingdom}\}$. Given the count matrix at a lower level, each element of the count matrix at the upper level can be calculated by $y_{ij}^{(l)}=\sum_{\{j':g_{jj'}=1\}} y_{ij'}^{(l-1)}$.

One property of the DM model is that if $(y_{1},\ldots,y_{p})\sim\text{DM}(\alpha_{1},\ldots,\alpha_{p})$, and if any two random variables, say $y_j$ and $y_{j'}$, are dropped from the vector and replaced by their sum $y_j+y_j'$, then we have $(y_{1},\ldots,y_j+y_{j'},\ldots,y_{p})\sim\text{DM}(\alpha_{1},\ldots,\alpha_{j}+\alpha_{j'},\ldots,\alpha_{p})$. This aggregation property can be used to derive the abundance matrices in different levels sequentially just from the one at the lowest available level via $\alpha_{ij}^{(l)}=\sum_{\{j':g_{jj'}=1\}} \alpha_{ij'}^{(l-1)}$. For a joint inference, we suggest the following: 1) fit the bottom-level model to the microbiome count matrix $\bm{Y}^{(1)}$, which is at level $1$, and infer the corresponding latent relative abundance matrix $\bm{A}^{(1)}$; 2) summarize the abundance matrices at each upper levels, $\bm{A}^{(2)},\ldots,\bm{A}^{(L)}$; and 3) fit the top-level model to all abundance matrices from level $1$ to $L$, independently.

For the ZINB model, the aggregation property does not hold. We assume that the size factor estimation should be irrelevant to the choices of microbiome count data at different taxonomic levels. Therefore, we consider the following scheme for a joint inference: 1) fit the bottom-level model to the microbiome count matrix $\bm{Y}^{(1)}$ and infer the corresponding latent relative abundance matrix $\bm{A}^{(1)}$, as well as the sample-specific size factor $\bm{s}$; 2) fit the bottom-level model to the microbiome count matrices at each upper levels with fixed $\bm{s}$, and obtain the corresponding latent relative abundance matrices $\bm{A}^{(2)},\ldots,\bm{A}^{(L)}$; and 3) fit the top-level model to all the abundance matrices from level $1$ to $L$, independently.

For both DM and ZINB models, the last implementation is to individually fit the top-level model to the abundance matrix at each taxonomic level, although some efforts could be made to sharpen the inference. One proposal is to replace the independent Bernoulli prior with a Markov random field (MRF) prior, which incorporates information from the taxonomic classification system, on the selection of discriminatory microbial features. This could encourage two connected taxa in the phylogenetic tree to be both selected. In particular, we consider the MRF prior on each $\gamma_j$ at level $l$ as
\begin{equation}
\label{gamma_prior}
p(\gamma_j^{(l)}|\bm{\gamma}^{(l-1)},\bm{\gamma}^{(l+1)})= \frac{\exp\left(\gamma_j^{(l)}\left(d+f\sum_{l'\in\{l-1,l+1\}}\sum_{j':g_{jj'}=1}\gamma_{j'}^{(l')}\right)\right)}{1+\exp\left(d+f\sum_{l'\in\{l-1,l+1\}}\sum_{j':g_{jj'}=1}\gamma_{j'}^{(l')}\right)},
\end{equation}
with hyperparameters $d$ and $f$ to be chosen. According to (\ref{gamma_prior}) those taxa that have a direct evolutionary relationship are more likely to be jointly selected. The hyperparameter $d$ controls the sparsity of the prior model, while $f$ affects the probability of selection of a feature according to the status of its connected taxa. 

\section{Model Fitting}\label{posterior}
In this section, we briefly describe the Markov chain Monte Carlo (MCMC) algorithm for posterior inference, while the detailed description is in Section S2 of the supplement. Our inferential strategy allows us to simultaneously infer the latent relative abundance of each taxon $j$ (at different taxonomic levels indexed by $l$) in each subject $i$, while identifying the discriminating taxa through $\bm{\gamma}^{(l)}$'s.

\subsection{MCMC algorithm}\label{MCMC}
Our primary interest lies in the identification of discriminating taxa via the selection vectors $\bm{\gamma}^{(l)}$'s, or $\bm{\gamma}$ if no phylogenetic tree available. To serve this purpose, a MCMC algorithm is designed based on Metropolis search variable selection algorithms \citep{George1997,Brown1998}. As discussed in Section \ref{Gaussian_mixture}, we have integrated out the mean and variance components in Equation (\ref{zinb_level2}). This step helps us speed up the MCMC convergence and improve the estimation of $\bm{\gamma}^{(l)}$'s. (More details are available in Section S2 in the supplement.)

\subsection{Posterior inference} 
An efficient summarization of $\bm{\gamma}^{(l)}$ is to select the taxa based on their marginal distributions. In particular, we estimate marginal posterior probabilities of inclusion (PPI) of a single taxon by $\text{PPI}^{(l)}_j=\sum_{b=1}^B\left( \gamma_j^{(l)} \text{ at iteration }b\right) / B$, where $B$ is the total number of iterations after burn-in. The marginal PPI represents the proportion of MCMC samples in which a taxon is selected to be discriminatory. A set of differentially abundant taxa can be picked based on their PPIs. For example, the selection can be done by including those taxa with marginal PPIs greater than a pre-specified value such as $0.5$. Alternatively, we can choose the threshold that controls for multiplicity \citep{Newton2004}, which guarantees the expected Bayesian false discovery rate (FDR) to be smaller than a number. The Bayesian FDR is calculated as follows,
\begin{align}\label{BFDR}
\text{FDR}(c_\gamma) = \frac{\sum_{l=1}^L\sum_{j=1}^{p}(1-\text{PPI}_j^{(l)})\text{I}(1-\text{PPI}_j^{(l)}<c_\gamma)}{\sum_{l=1}^L\sum_{j=1}^{p}\text{I}(1-\text{PPI}_j^{(l)}<c_\gamma)}.
\end{align}
Here $c_\gamma$ is the desired significance level, with $c_\gamma=0.05$ being generally used in other parametric/nonparametric test settings for microbiome studies.

\section{Simulation}\label{simulation}
In this section, we briefly summarize the simulation studies. The detailed description is available in Section \ref{S_simulation} in the supplement. 

We use both simulated and synthetic data to show that the proposed Bayesian framework generally outperforms alternative methods currently used in the field of microbial differential abundance analysis, which include: 1) Analysis of variance (ANOVA); 2) Kruskal-Wallis (KW) test; 3) \texttt{edgeR} \citep{robinson2010edger}; 4) \texttt{DESeq2} \citep{love2014moderated}; and 5) \texttt{metagenomeSeq} \citep{Paulson2013}. The first two are parametric/nonparametric methods for testing whether samples originate from the same distribution. The third and fourth are representative methods for analyzing RNA-Seq count data. The last one, \texttt{metagenomeSeq}, assumes a zero inflated Gaussian model on the log-transformed counts, and performs a multiple groups test on moderated F-statistics. 

We simulated species-level datasets with $n=24$ or $108$ samples, and $p=1,000$ features, $50$ of which were truly discriminatory among $K=2$ or $3$ groups. The hierarchical formulations of the generative models are discussed in Section S3.1 and presented in Table S1 in the supplement. The prior specifications are presented in Section S3.2, and a follow-up sensitivity analysis is shown in Section S3.5 in the supplement. To quantify the accuracy of identifying discriminatory features via the binary vector $\bm{\gamma}$, we used two well-accepted measures of the quality of binary classifications: 1) area under the curve (AUC) of the receiver operating characteristic (ROC); and 2) Matthews correlation coefficient (MCC) \citep{matthews1975comparison}, which is defined in Section S3.3 in the supplement.

With the results from this simulation studies summarized in Figure \ref{simu_fig_1} and \ref{simu_fig_2}, Table \ref{simuauc} - \ref{synth} in the supplement, we concluded as follows: 1) the DPP normalization method in our Bayesian ZINB model showed advantages of making unbiased estimation on the size factors $\bm{s}$, and outputting their uncertainty; 2) Our Bayesian ZINB model with DPP on $\bm{s}$ always achieved the highest performance in terms of AUC and MCC, and our Bayesian DM model maintained the second-best in general, even if the generative schemes of some simulated datasets were not favor the fitted model; 3) Decreasing either the sample size $n$ or the effect size would lead to greater disparity between ours and the others.

\section{Real Data Analysis}\label{realdata}
\subsection{Colorectal cancer study}\label{realdata_1}
Colorectal cancer (CRC) is the third most common cancer diagnosed in men and women in the United States \citep{arnold2017global}. There have been an increasing number of studies suggesting an association between CRC and the gut microbiome \citep{sears2014microbes}. We applied our model to a colorectal cancer gut microbiome dataset published by \cite{zeller2014potential}. The cohort consisted of 199 individuals from Europe (91 CRC patients and 108 non-CRC controls) and the disease status was confirmed by intestinal biopsy. The original metagenomic sequence data from the fecal samples were available from the
European Nucleotide Archive database (accession number ERP005534). We used %MetaPhlAn2 \citep{truong2015metaphlan2} and 
curatedMetagenomicData \citep{pasolli2017accessible} to obtain the taxonomic abundance table of 199 patients with 3940 detected taxa. After the quality control step (details in Section S4.1 in the supplement), we were left with $n = 182$ patients and $p = 492$ taxa in total.

We applied the proposed ZINB-DPP model to detect the differentially abundant taxa.  We chose weakly informative priors as discussed in Section S3.2 in the supplement. Specifically, we set the shape parameters $a_0=a_1=\ldots=a_k = 2$ and the scale parameters $b_0=b_1=\ldots=b_k=1$ for variance components $\sigma_{0j}^2$ and $\sigma_{kj}^2$. Next, we let $h_0 = h_1 = \ldots = h_K = 50$. Our sensitivity analysis (presented in Section S3.5 in the supplement) shows the posterior inference on $\bm{\gamma}$ remained almost the same when those values were in the range of $10$ to $100$. As indicated by \cite{stingo2013integrative}, larger values of these hyperparameters would encourage the selection of only very large effects whereas smaller values would encourage the selection of smaller effects. We further set $d=-2.2$ and $f=0.5$ as the default choice of the MRF prior. It means that if a taxon does not have any neighbor as a discriminatory taxon, its prior probability that it has differential abundance equals to $\exp(-2.2)/(1+\exp(-2.2))=0.1$. Finally, we specified $M = n/2, ~c_s = 0, ~\sigma_s = 1, ~\tau_{\eta} = 1, ~a_t = b_t = 1, ~a_m = b_m = 1,~a_{\phi} = b_{\phi} = 0.001$ and $a_{\pi} = b_{\pi} = 1$. Our inference used four independent MCMC chains with 20,000 iterations each (first 10,000 iterations for burn-in, and the rest for inference). We calculated the PPIs for all chains and found their pairwise correlation coefficients range from 0.954 to 0.964, which suggested good MCMC convergence. We then averaged the outputs of all chains as final results and selected the discriminating taxa by contrlloing the Bayesian FDR at $1\%$ level.

\begin{figure}[!h]
	\centering
	\includegraphics[width=1.0\textwidth]{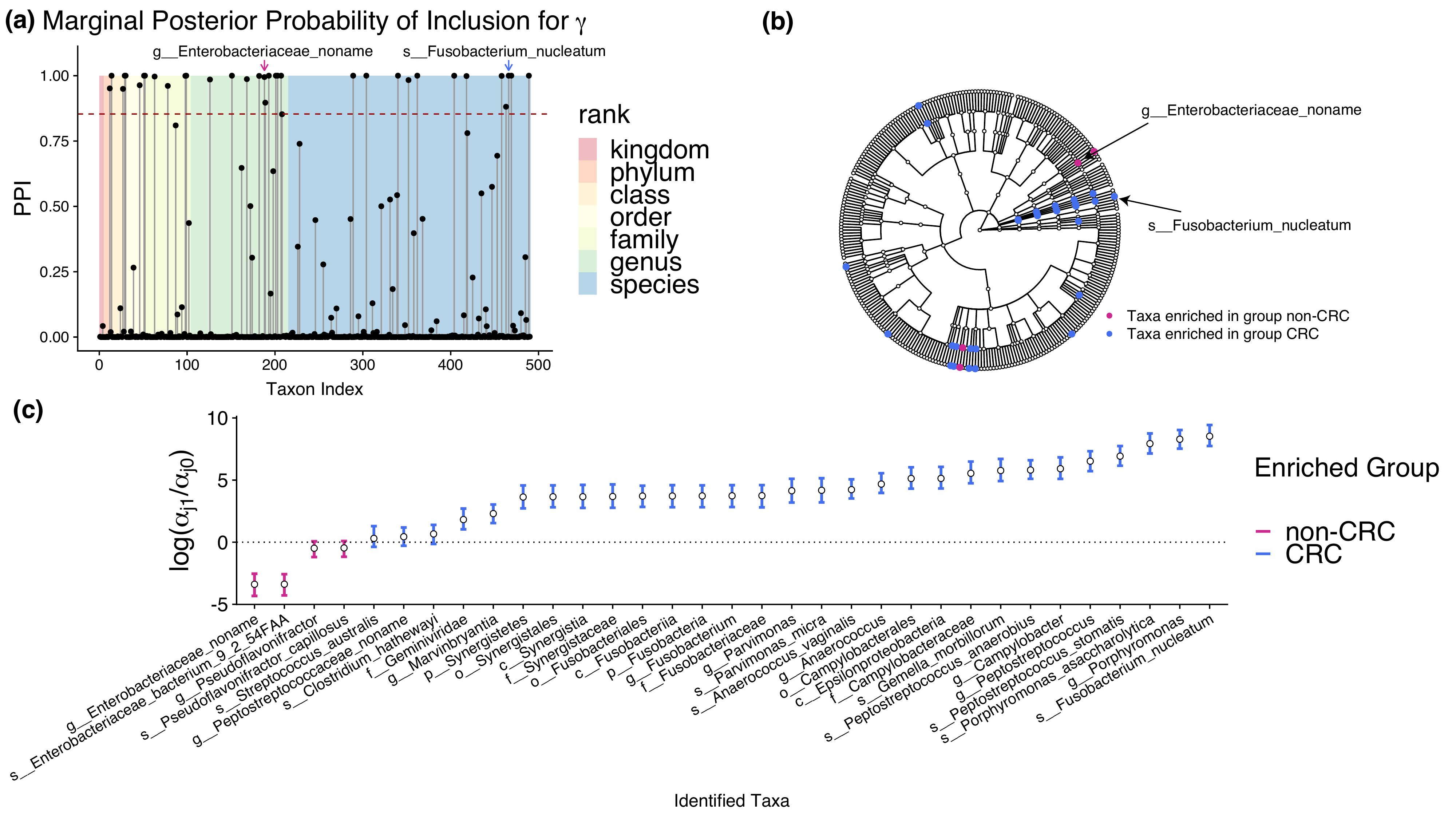} 
	\caption[]{Colorectal cancer study study: (a) plot of $\gamma$ PPIs with the horizontal dashed line representing the threshold controlling the Bayesian false discovery rate of 1\%; (b) cladogram of the identified discriminating taxa (shown in dots) with each arrow pointing out the taxon with largest absolute values of $\log(\alpha_{1j}/\alpha_{0j})$ in one patient group; (c) 95\% credible intervals for $\log(\alpha_{j1} / \alpha_{j0})$ of the reported discriminating taxa.}
	\label{fig:realdata1}
\end{figure}

In total, the ZINB-DPP model detected 33 differentially abundant taxa (see Figure \ref{fig:realdata1}(a)). Among them, \textit{Fusobacterium nucleatum} (\textit{Fn}) had the largest PPI value and the largest effect size (see Figure \ref{fig:realdata1}(c)). \textit{Fn} is a well-known taxon associated with CRC as reported by a series of studies. \cite{castellarin2012fusobacterium} observed that the over-abundance of \textit{Fn} was associated with CRC tumor specimens, and they suggested that \textit{Fn} can invade colonic mucosa and thus induces local inflammations. Later, \cite{kostic2013fusobacterium} and \cite{rubinstein2013fusobacterium} confirmed the causative role of \textit{Fn}, and they experimentally showed that \textit{Fn} invasion would replenish tumor-infiltrating immune cells and generate a tumorigenic microenvironment to promote colorectal neoplasia. At the species level, our model additionally detected \textit{Peptostreptococcus stomatis} and \textit{Porphyromonas asaccharolytica}. These species were supported from biological literatures (Table \ref{species}). Interestingly, all the above taxa were also reported to be the most important predictors in a prediction model to detect CRC \citep{zeller2014potential}. Besides the discovery of CRC-enriched taxa, our model also reported the taxa enriched in healthy controls and depleted in CRC patients. For example, \cite{marchesi2011towards} found that \textit{Eubacteriaceae} was underrepresented in CRC tissue, and our model detected the corresponding genus and species. Similarly, \cite{warren2013co} found \textit{Pseudoflavonifractor} was over-represented in control samples in a CRC study. %\textit{Ruminococcus champanellensis}, an obligate anaerobic bacteria, was a newly found species and less studied. However, its genus \textit{Ruminococcus} was found to be decreased in colorectal cancer \citep{brennan2016gut}. Similarly, \textit{Eubacteriaceae} was found to be underrepresented in CRC tissue \cite{marchesi2011towards}. 
In all, 11 differentially abundant species were identified by our model, 6 of which have previously been reported as potentially important in CRC pathophysiology based on existing biology literatures (summarized in the last column of Table \ref{species}).

We visualized all differentially abundant taxa in a cladogram (see Figure \ref{fig:realdata1}(b)). These taxa were clustered along the branches of the phylogenetic tree. This phenomenen was modeled by our MRF prior and could guide biological studies. For example, the branch of \textit{Fn} were from phylum level (\textit{Fusobacteria}) to species level, which suggested that the shared sequence similarities in the \textit{Fusobacteria} branch were positively associated with CRC. Our model also detected a branch of gram-negative bacteria, from genus level (\textit{Campylobacter}) to class level (\textit{Epsilonproteobacteria}). Utilizing the phylogenetic tree structure, biologists can select bacteria species under the genus level of \textit{Fusobacterium} and \textit{Campylobacter} in validations experiment. In this direction, \cite{warren2013co} reported significant co-occurrence of \textit{Fusobacterium} and \textit{Campylobacter} species observed in individual CRC tumors.

\begin{figure}[!h]
	\centering
	\includegraphics[width = 0.7\linewidth]{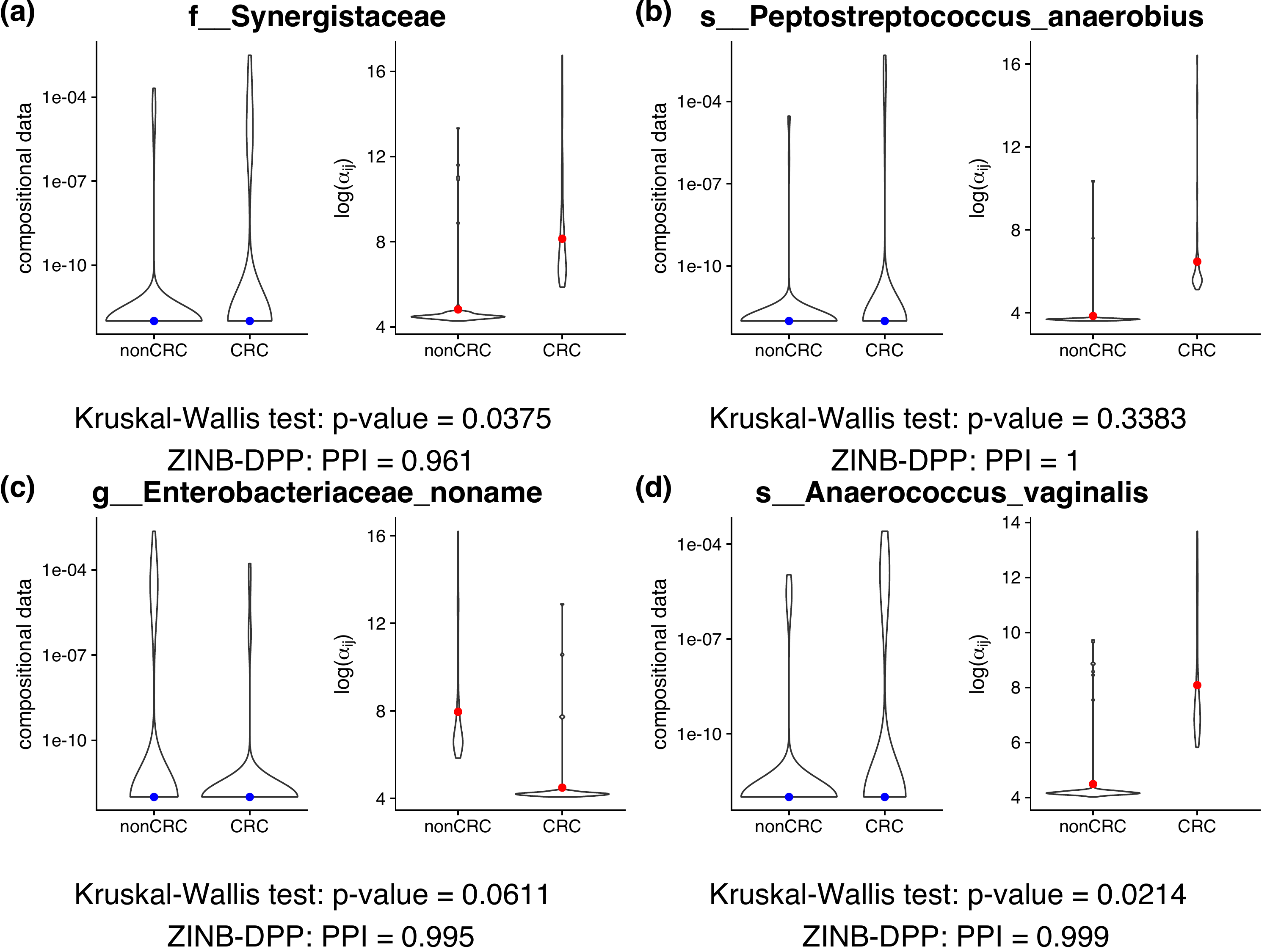}
	\caption{Colorectal cancer study: violin plots comparing the normalized abundance for four taxa detected by ZINB-DPP model but missed by the Kruskal-Wallis (KW) test. For each case, the left part compares the compositional data and the right part compares the latent elative abundance ($\alpha_{ij}$) on the log scale. The blue dots represent the group median and the red dots represent the group mean.}
	\label{casedist}
\end{figure}

\begin{figure}[!h]
	\centering
	\includegraphics[width=.8\textwidth]{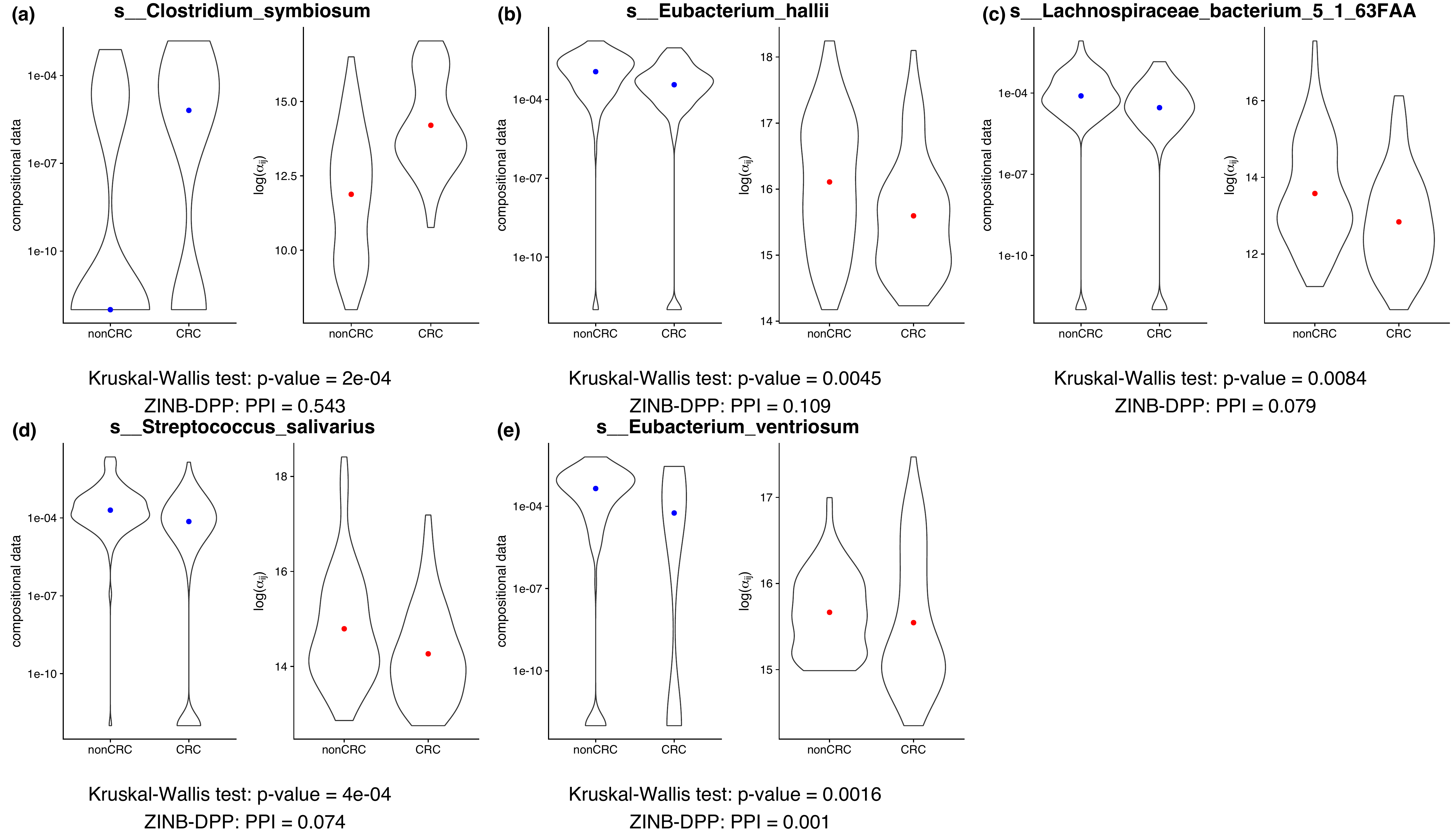} 
	\caption{Colorectal cancer study: violin plots comparing the normalized abundance for all the five species detected by the Kruskal-Wallis (KW) test but missed by the ZINB-DPP model. For each case, the left part compares the compositional data and the right part compares the relative abundance ($\alpha_{ij}$) on the log scale. The blue dots represent the group median and the red dots represent the group mean. }
	\label{dist_compare}
\end{figure}

We evaluated alternative approaches including ANOVA, KW test, \texttt{DESeq2}, \texttt{edgeR} and \texttt{MetagenomeSeq}, and compared their analysis results with our models. As the KW test was widely used in comparative metagenomic data analysis, we compared the detected differentially abundant taxa by KW to those by ZINB-DPP. For other methods, we presented the comparisons in the supplement (see Figure \ref{tree_other_1}). The KW test reported 30 taxa under the 1\% significance level threshold on the BH adjusted p-values, and 19 of them were also found by the ZINB-DPP model. For the taxa that were detected by only one but not the other, we examined their the actual data distributions and biological literatures. Figure \ref{casedist} compared the distributions of four out of 14 taxa detected by ZINB-DPP model but missed by the KW test (two species (b) and (d) were also listed in Table \ref{species}). We illustrated the taxonomic compositions in logarithm scale as KW compared the group medians, and the latent relative abundances ($\alpha_{ij}$) in logarithm scale for ZINB-DPP. 
Notably, there were visible separations of latent relative abundances only from ZINB-DPP, as our model can properly adjust for sample heterogeneity and zero pattern (i.e. true zero or missing). 
In addition, the species-level taxa \textit{Peptostreptococcus anaerobius} and family-level taxa \textit{Synergistaceae} (Figure \ref{casedist}(a) and \ref{casedist}(b)) were supported by the recent literature \citep{tsoi2017peptostreptococcus,coker2019enteric}, while the bottom two were the novel findings by ZINB-DPP model with biological evidence on their higher taxonomic levels.
On the contrary, as a nonparametric test comparing the group median (represented by blue dots) of the compositional taxa data, KW failed to distinguish different medians under a 1\% significance level (Figure \ref{casedist}). Moreover, compared to the bi-level design of our proposed statistical framework, KW test cannot estimate the effect sizes of the detected taxa (Figure \ref{fig:realdata1}(c)) and that poses extra challenges to interpret the effect directions and sizes for biologists. 
Aside of the taxa only detected by ZINB-DPP, we also examined those missed by ZINB-DPP. Figure \ref{dist_compare} compared the distributions of all five taxa that were given by the KW test but not ZINB-DPP, and Table \ref{species2} showed the adjusted p-values and PPIs from both methods.
For species \textit{Clostridium symbiosum}, KW had a small p-value that was likely driven by the different fractions of zeros between groups ($65\%$ of the counts in the non-CRC group were zeros while $31\%$ in the CRC group). For the rest four species, the violin plots showed both similar medians (represented by the blue dots) for the compositional data and similar means (represented by the red dots) for the latent relative abundances inferred by the ZINB-DPP. Although adjusted p-values from KW test were all significant at the $0.01$ level, there is a lack of clear patterns of separation between groups.

\subsection{Schizophrenia study}\label{realdata_2}
Schizophrenia is a life-threatening neuropsychiatric disorder typically manifesting as hallucinations, delusions, and social withdrawal. The study of gut-brain axis suggested that the microbiota can affect psychiatric symptomatology \citep{fond2015psychomicrobiotic}, and is a key component in diseases related to neurodevelopment and stress responses \citep{rea2016microbiome}. In order to evaluate the proposed model in a dataset with smaller sample size, here we analyzed the metagenome-sequenced oropharynx samples of 16 schizophrenia patients and 16 controls in a study by \cite{castro2015composition}. We processed the sequence data, implemented the quality control steps, and obtained the taxonomic abundance matrix of $n = 27$ samples and $ p = 271 $ taxa (see details in Section S4.1 in the supplement).

\begin{figure}[!h]
	\centering
	\includegraphics[width=1.0\textwidth]{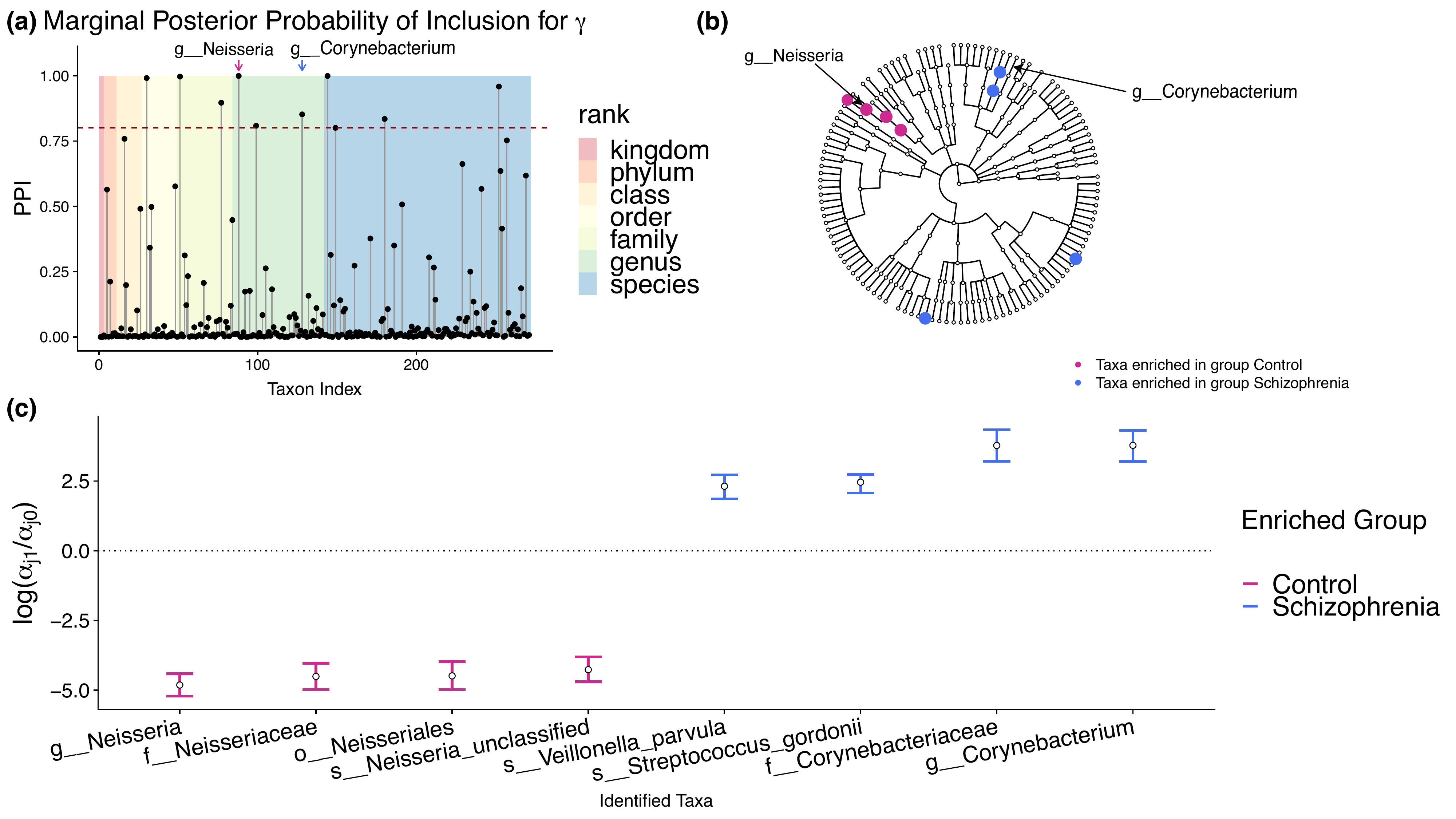} 
	\caption[]{(a) plot of $\gamma$ PPI with the horizontal dashed line representing the threshold controlling the Bayesian false discovery rate of 5\%; (b) cladogram of the identified discriminating taxa (shown in dots) with each arrow pointing out the taxon with the largest absolute values of $\log(\alpha_{1j}/\alpha_{0j})$ in one patient group; (c) 95\% credible intervals for $\log(\alpha_{j1} / \alpha_{j0})$ of the reported discriminating taxa.}
	\label{fig:realdata2}
\end{figure}

We used the ZINB-DPP model with the default priors  (described in Section S3.2 in the supplement) to analyze this dataset. Four independent chains were run with randomly initialized starting points. After discarding the first half of 20,000 iterations for each chain, we calculated pairwise correlation coefficients of PPIs (ranging from 0.978 to 0.987), which indicated that the MCMC chains were convergent. Figure \ref{fig:realdata2} presented the identified differentially abundant taxa, their phylogenetic relationships and estimated effect sizes. The top two taxa with the largest effect sizes belonged to family \textit{Corynebacteriaceae}, suggesting their damaging role to the schizophrenia. A relevant study by \cite{strati2017new} reported that the abundance of \textit{Corynebacterium} was significantly increased in a cohort with autism spectrum disorder (a neurodevelopmental disorder). \cite{bavaro2011pentapeptide} reported that a species from \textit{Corynebacterium} was associated with the human neural protein network. Both studies hinted a strong associative effect of \textit{Corynebacterium} in mental disorders. Our model additionally detected \textit{Veillonellaceae}. Significant alternation of \textit{Veillonellaceae} level was observed between healthy people and patients in various studies of nervous system disorders, such as autism, gastrointestinal disturbances, etc \citep{kelly2017cross}. Besides case-enriched taxa, our model also identified four control-enriched taxa under the order \textit{Neisseriales}. Several taxa from this phylogenetic tree branch have been reported to be associated with psycological diseases. For example, \cite{prehn2018reduced} detected the altered levels of \textit{Neisseriaceae} and \textit{Neisseria sp.} for patients with a specific type of psychosocial and behavioral disease. We compared with other alternative models and most of them also detected \textit{Neisseria} and \textit{Neisseria sp.} to be differentially abundant (see details in Figure \ref{tree_other_2} in the supplement). In addition, our model identified \textit{Streptococcus gordonii} to be significantly enriched in schizophrenia patients. This species had not been widely investigated in psychiatric studies, and the adjusted p-value from the Kruskal–Wallis test was above a significance level of $0.05$. Given the small sample size ($n = 27$) and  frequent zero-data ($10$ out of $27$), these results must be interpreted with caution. Ultimately, additional studies which either corroborate clinical association or intimate causality in a preclinical model would be merited.

%A recent study of attention-deficit/hyperactivity disorder (ADHD) showed that patients with ADHD had elevated levels of \textit{Neisseriaceae} and \textit{Neisseria spec} \citep{prehn2018reduced}. ADHD is a psychiatric disorder characterized by hyperactivity, impulsivity and attention problems. Interestingly, \cite{hamshere2013shared} identified significant shared genetic susceptibility between childhood ADHD and schizophrenia.
%the adjusted $p$-value from the Kruskal–Wallis test is above a significance level of 0.01. Given the small sample size ($n = 27$) and along with the presence of too many zeros (10 out of 27), undermines the reliability of the test results. As suggested by our model, biological validations would be a promising next step for better understanding the schizophrenia pathogenesis. 

\section{Conclusion}\label{conclusion}
In this paper, we have proposed a Bayesian hierarchical framework for analyzing microbiome sequencing data. Our bi-level framework offers flexibility to choose different normalization models and differential abundance analysis models, in distinct levels. Under this framework, we showed that our Bayesian nonparametric prior with stochastic constraints can reduce estimation bias and improve the posterior inferences of the other parameters of interests. Notably, our application of the Dirichlet process prior is not restricted to microbiome data analysis, and it is generally applicable to other types of heterogeneous sequence data \citep{Li2017}. Moreover, our model can jointly analyze multiple microbes at different taxonomic levels while offering well-controlled Bayesian false discovery rates. In addition, our model is applicable for studies with more than 2 disease outcomes, such as multiple patient subtypes. Additionally, our model can support the detection of discriminating taxa among all patient subgroups or between any pair of them. The MCMC algorithm is implemented using the \texttt{R} package \texttt{Rcpp} to improve computational efficiency. The code used in all simulation and real data analyses are available upon request. 

As a summary of model performance, the ZINB-DPP model consistently outperforms commonly used methods in model-based simulations, synthetic data simulations and two real data analyses. The advantages become more obvious as either the sample size or the effect size decreases. In two case studies, our results are consistent with the current biological literatures. For researchers interested in more performance details, we have evaluated other competing methods on the same datasets (CRC study and schizophrenia study) and have presented the results in the supplement. Specifically, we noticed that the sparsity observed in microbiome data could impair the statistical power of ANOVA. Meanwhile, \texttt{edgeR} and \texttt{DESeq2} tend to have higher false positive rates, whereas \texttt{metagenomeSeq} produces relatively conservative results compared to our model. These findings are consistent with \citep{weiss2017normalization} and are helpful to future microbiome data analysis.
% based on detected species. Along this direction, it is not suitable to compare our model to the methods for detecting differential microbiome communities (jointly testing a group of taxa). 
%% the last sentense %%
Our model framework can be naturally extended to other analysis scenarios. For example, the inferred latent abundance can be treated within a sample normalized distribution. It is thus applicable to longitudinal analysis, which can capture the dynamic structure in microbiome studies; or to differential network analysis, which can investigate the complex interactions among microbial taxa. In all, the proposed Bayesian framework provides more powerful microbiome differential abundance analyses and is suitable for multiple types of microbiome data analysis.

%\begin{sidewaystable}
\begin{table}[h]
	\centering
	\caption{A list of multivariate count generating processes and their characterizations}\label{Table 1}
	\footnotesize
	\begin{tabular}{@{}|l|l|l|c|c|c|l|@{}}
		\hline
		%& & Uneven & Zero- & Over- & \\
		& $\mathcal{M}(\bm{y}_{i\cdot};\bm{\alpha}_{i\cdot},\bm{\Theta})$ & $\bm{\Theta}$& \rotatebox{90}{Uneven depth } & \rotatebox{90}{Zero-inflation } & \rotatebox{90}{Over-dispersion } & Example\\\hline
		%log & $\mathbb{I}_0\left(\alpha_{i1}=\log{(y_{i1}+1)},\ldots,\alpha_{ip}=\log{(y_{ip}+1)}\right)$ &  & &  &\\\hline
		%Frac. & $\mathbb{I}\left(\alpha_{i1}=y_{i1}/Y_{i\cdot},\ldots,\alpha_{ip}=y_{ip}/Y_{i\cdot}\right)$ &$\bullet$ & & &\\\hline
		Multi& $\text{Multi}(\bm{y}_{i\cdot};Y_{i\cdot},\alpha_{i1},\ldots,\alpha_{ip})$ &&$\bullet$ & & &\\\hline
		DM & $\text{DM}(\bm{y}_{i\cdot};\alpha_{i1},\ldots,\alpha_{ip})$ &&$\bullet$ & & $\bullet$ & \cite{La2012} \\\hline
		%Gaussian & $\prod_{j=1}^p\exp\left(\text{Normal}\left(s_i\alpha_{ij},\sigma_{j}^2\right)\right)-1$ &$\bullet$ & & $\bullet$ &\\\hline
		Poisson & $\prod_{j=1}^p\text{Poi}(y_{ij};s_i\alpha_{ij})$ & $\{\bm{s}\}$ &$\bullet$ & & & \cite{brown2011gut} \\\hline
		NB & $\prod_{j=1}^p\text{NB}(y_{ij};s_i\alpha_{ij},\phi_{j})$ & $\{\bm{s},\bm{\phi}\}$ &$\bullet$ & & $\bullet$ & \cite{zhang2017negative}  \\\hline
		ZIG & $\prod_{j=1}^p\pi_i(Y_{i\cdot})\text{I}(y_{ij}=0)+$ & $\{\bm{\sigma},\bm{\pi}\}$ &$\bullet$ & $\bullet$ & $\bullet$ & \cite{Paulson2013}\\
		& $\quad(1-\pi_i(Y_{i\cdot}))\text{N}\left(\log(y_{ij}+1);\alpha_{j},\sigma_{j}^2\right)$ && & & &\\\hline
		ZIP& $\prod_{j=1}^p\pi_i\text{I}(y_{ij}=0)+$ & $\{\bm{s},\bm{\pi}\}$  &$\bullet$ & $\bullet$ & & \cite{cheung2002zero}  \\
		& $\quad(1-\pi_i)\text{Poi}(y_{ij};s_i\alpha_{ij})$ &&&  & & \\\hline
		ZINB & $\prod_{j=1}^p\pi_i\text{I}(y_{ij}=0)+$ & $\{\bm{s},\bm{\phi},\bm{\pi}\}$ &$\bullet$ & $\bullet$ & $\bullet$ & \cite{fang2016zero} \\
		& $\quad(1-\pi_i)\text{NB}(y_{ij};s_i\alpha_{ij},\phi_{j})$ &&& & &\\\hline
		\multicolumn{7}{p{0.95\textwidth}}{Abbreviations: Multinomial (Multi); Dirichlet-multinomial (DM); Negative binomial (NB); Zero-inflated Gaussian (ZIG); Zero-inflated Poisson (ZIP); Zero-inflated negative binomial (ZINB).}
	\end{tabular}
\end{table} 
%\end{sidewaystable}

%\begin{sidewaystable}
\begin{table}[h]
	\centering
	\caption{List of commonly used normalization techniques for sequencing count data}\label{Table 2}
	\footnotesize
	\begin{tabular}{@{}|l|l|l|l|@{}}
		\hline
		& Definition & Constraint & Reference\\\hline
		TSS & $\hat{s}_i\propto Y_{i\cdot}$ & $\sum_{i=1}^n\log s_i=0$ & \\\hline
		$^1$Q75 & $\hat{s}_i\propto q_i^{0.75p},$ & $\sum_{i=1}^n\log s_i=0$ & \cite{Bullard2010} \\\hline
		RLE & $\hat{s}_i\propto\text{median}_j\left\{y_{ij}/\sqrt[n]{\prod_{i'=1}^ny_{i'j}}\right\}$ & $\sum_{i=1}^n\log s_i=0$ & \cite{Anders2010} \\\hline
		$^2$TMM &
		$\hat{s}_i\propto\sum_{j=1}^py_{ij}\cdot\exp\left( \frac{\sum_{j\in G^*}\psi_j(i,r)M_j(i,r)}{\sum_{j\in G^*}\psi_j(i,r)}
		\right)$ & $\sum_{i=1}^n\log s_i=0$ & \cite{Robinson2010}\\\hline
		$^1$CSS  & $\hat{s}_i\propto\sum_{j=1}^py_{ij}\cdot\text{I}(y_{ij}\le q_i^{0.5p})$ & $\sum_{i=1}^n\log s_i=0$ & \cite{Paulson2013}\\\hline
		DPP & 
		$p(\log s_i|\cdot)=\sum_{m=1}^M\psi_m\Big[ t_m\text{N}(\nu_m,\sigma_s^2)+$
		& $\text{E}(\log s_i)=0$ & \cite{Li2017}\\
		& $\quad(1-t_m)\text{N}\left(\frac{c_s-t_m\nu_m}{1-t_m},\sigma_s^2\right)\Big]$ &  & \\\hline
		\multicolumn{4}{p{1.0\textwidth}}{Abbreviations: TSS is total sum scaling, Q75 is upper-quartile (i.e. $75\%$), RLE is relative log expression, TMM is trimmed mean by M-values, CSS is cumulative sum scaling, and DPP is Dirichlet process prior.}\\
		\multicolumn{4}{p{1.0\textwidth}}{$^1$Note for Q75 and CSS: $q_i^l$ is defined as the $l$-th quantile of all the counts in sample $i$, i.e. there are $l$ features in sample $i$ whose values $y_{ij}$'s are less than $q_i^l$.}\\
		\multicolumn{4}{p{1.0\textwidth}}{$^2$Note for TMM: the M-value $M_j(i,r)=\log(y_{ij}/Y_{i\cdot})/\log(y_{rj}/Y_{r\cdot})$ is the log-ratio of scaled counts between sample $i$ and the reference sample $r$, if not within the upper and lower $30\%$ of all the $M$-values (as well as the upper and lower $5\%$ of all the $A$-values, defined as $A_j(i,r)=\log\sqrt{y_{ij}/Y_{i\cdot}\cdot y_{rj}/Y_{r\cdot}}$, and the corresponding weight $\psi_{j'}(i,r)$ is the inverse of the approximate asymptotic variances, calculated as $\frac{Y_{i\cdot}-y_{ij'}}{y_{ij'}Y_{i\cdot}}+\frac{Y_{r\cdot}}{y_{rj'}Y_{r\cdot}}$ by the delta method.}
	\end{tabular}
\end{table}
%\end{sidewaystable} 

\begin{table}[!h]
	\centering
	\resizebox{0.9\columnwidth}{!}{%
		\begin{tabular}{l |T{5cm}| T{5cm}| c | l}
			\hline
			\hline
			\makecell{Species Name} &
			\makecell{ZINB-DPP \\ (with PPI)} & 
			\makecell{Kruskal-Wallis Test \\ (with adjusted p-value)} &
			\makecell{Figure Label} & \makecell{Evidence} \\
			\hline
			\multirow{2}{*}{\textit{Fusobacterium nucleatum}}& \multirow{2}{*}{\textbf{\underline{1.000} }}  & \multirow{2}{*}{ \underline{\bm{$<0.001$}}} & & \cite{castellarin2012fusobacterium,kostic2013fusobacterium} \\
			&  &   & &\cite{rubinstein2013fusobacterium}  \\
			\hline
			\textit{Clostridium hathewayi} & \textbf{\underline{1.000} } & \textbf{\underline{0.001}} & &\\  \hline
			\textit{Gemella morbillorum} & \textbf{\underline{1.000} } &  \underline{\bm{$<0.001$}} &  &\cite{kwong2018association}\\   \hline
			
			\textit{Peptostreptococcus stomatis} & \textbf{\underline{1.000} }  &   \underline{\bm{$<0.001$}} &  & \cite{purcell2017distinct, drewes2017high}\\   \hline
			\textit{Peptostreptococcus anaerobius} & \textbf{\underline{1.000} } & 0.338& Figure \ref{casedist}(b) &
			\cite{tsoi2017peptostreptococcus} \\   \hline
			\textit{Porphyromonas asaccharolytica} & \textbf{\underline{1.000} } &  \underline{\bm{$<0.001$}}& & \cite{flynn2016metabolic}\\   \hline
			\textit{Streptococcus australis} & \textbf{\underline{1.000} } &   \textbf{\underline{0.004}}&  &\\    \hline
			\textit{Anaerococcus vaginalis} &  \textbf{\underline{0.999}} &  0.021 & Figure \ref{casedist}(d) &\\   \hline
			\textit{Enterobacteriaceae bacterium} 9-2-54FAA &  \textbf{\underline{0.999}} &  0.061& &\\  \hline 
			\textit{Pseudoflavonifractor capillosus} &  \textbf{\underline{0.999}} & 0.032 & &\\  \hline
			\textit{Parvimonas micra}&  \textbf{\underline{0.983}} &  \textbf{\underline{0.001}} & & \cite{purcell2017distinct, drewes2017high} \\  \hline
			\hline
		\end{tabular}
	}
	\caption{Colorectal cancer study: the species level detections from ZINB-DPP model under the Bayesian FDR of $1\%$. The corresponding adjusted p-values from the Kruskal–Wallis test are also supplied. In each row, an underlined posterior probability of inclusion (PPI) or adjusted p-value (under a significance level of $1\%$) means the species is selected as differentially abundant between two groups by the corresponding method. Column ``Figure Label" indicates the figure that compares the distribution of specific species. Column ``Evidence" lists the relevant literatures supporting the selection for the species.}
	\label{species}
\end{table}

\begin{table}[!h]
	\centering
	\resizebox{0.9\columnwidth}{!}{%
		\begin{tabular}{l |T{5cm}| T{5cm}| c }
			\hline
			\hline
			\makecell{Species Name} &
			\makecell{ZINB-DPP \\ (with PPI)} & 
			\makecell{Kruskal-Wallis Test \\ (with adjusted p-value)} &
			\makecell{Figure Label}  \\
			\hline
			\textit{Clostridium symbiosum} & 0.543 &  \underline{\bm{$<0.001$}}&  Figure \ref{dist_compare}(a) \\ \hline
			\textit{Eubacterium hallii} & 0.109 &\textbf{\underline{0.004}} &  Figure \ref{dist_compare}(b)\\  \hline
			\textit{Lachnospiraceae bacterium} 5-1-63FAA & 0.079 &  \textbf{\underline{0.008}} &  Figure \ref{dist_compare}(c)\\  \hline
			\textit{Streptococcus salivarius} & 0.074 &  \underline{\bm{$<0.001$}} &  Figure \ref{dist_compare}(d)\\  \hline
			\textit{Eubacterium ventriosum} & 0.001 &  \textbf{\underline{0.002}}&  Figure \ref{dist_compare}(e) \\  \hline
			\hline
		\end{tabular}
	}
	\caption{Colorectal cancer study: species level detections by the Kruskal–Wallis test but not the ZINB-DPP model under the Bayesian FDR or the significance level of $1\%$. Column ``Figure Label" indicates the figure that compares the distribution of the five species. }
	\label{species2}
\end{table}

\clearpage
\section*{S1 \quad Graphical Formulations of the Proposed Models}

\begin{figure}[!h]
	\centering
	\begin{minipage}[t]{.6\textwidth}
		\centering
		\subfloat[]{\includegraphics[width=\linewidth]{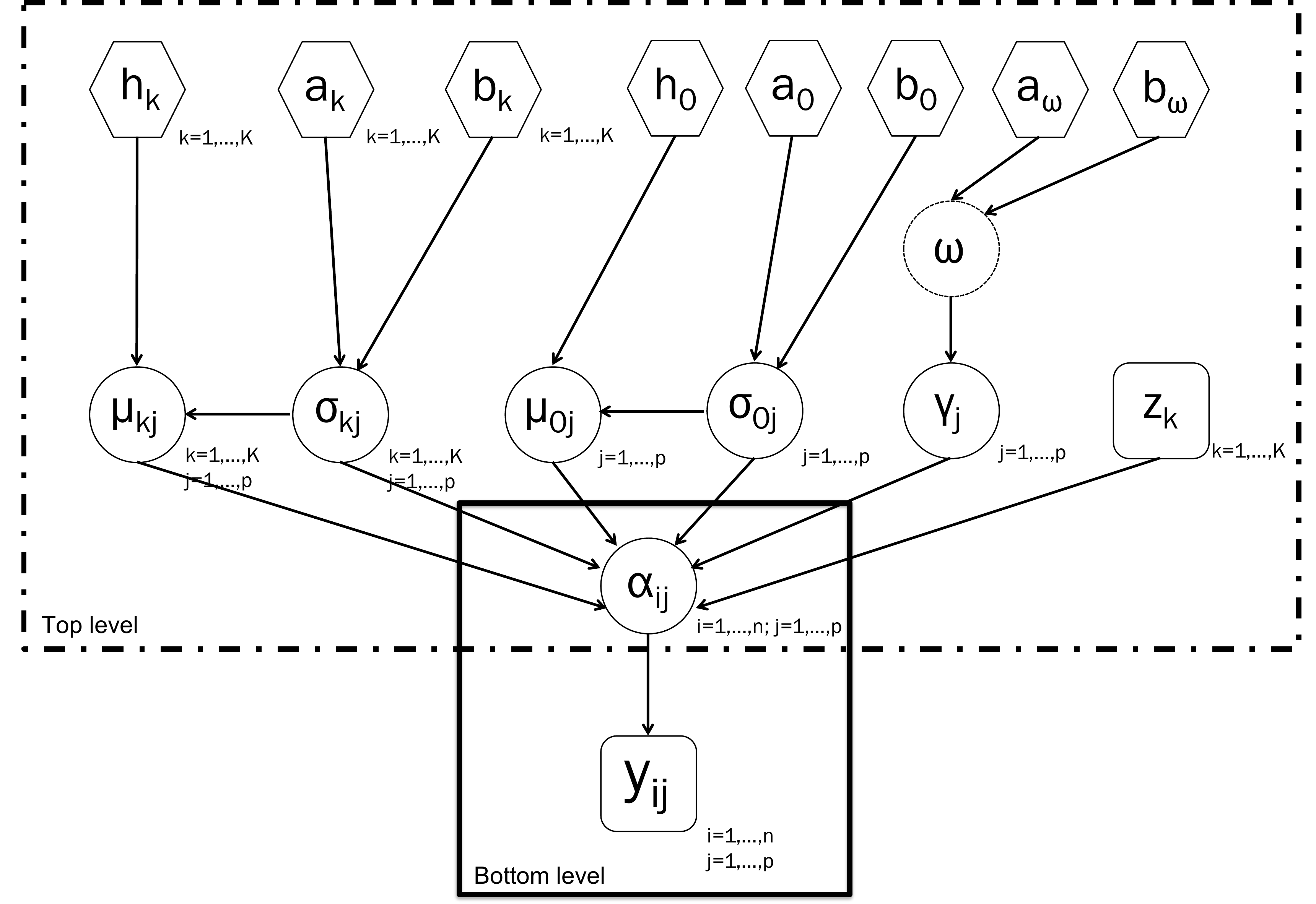}}
	\end{minipage}\\
	\begin{minipage}[t]{.8\textwidth}
		\centering
		\subfloat[]{\includegraphics[width=\linewidth]{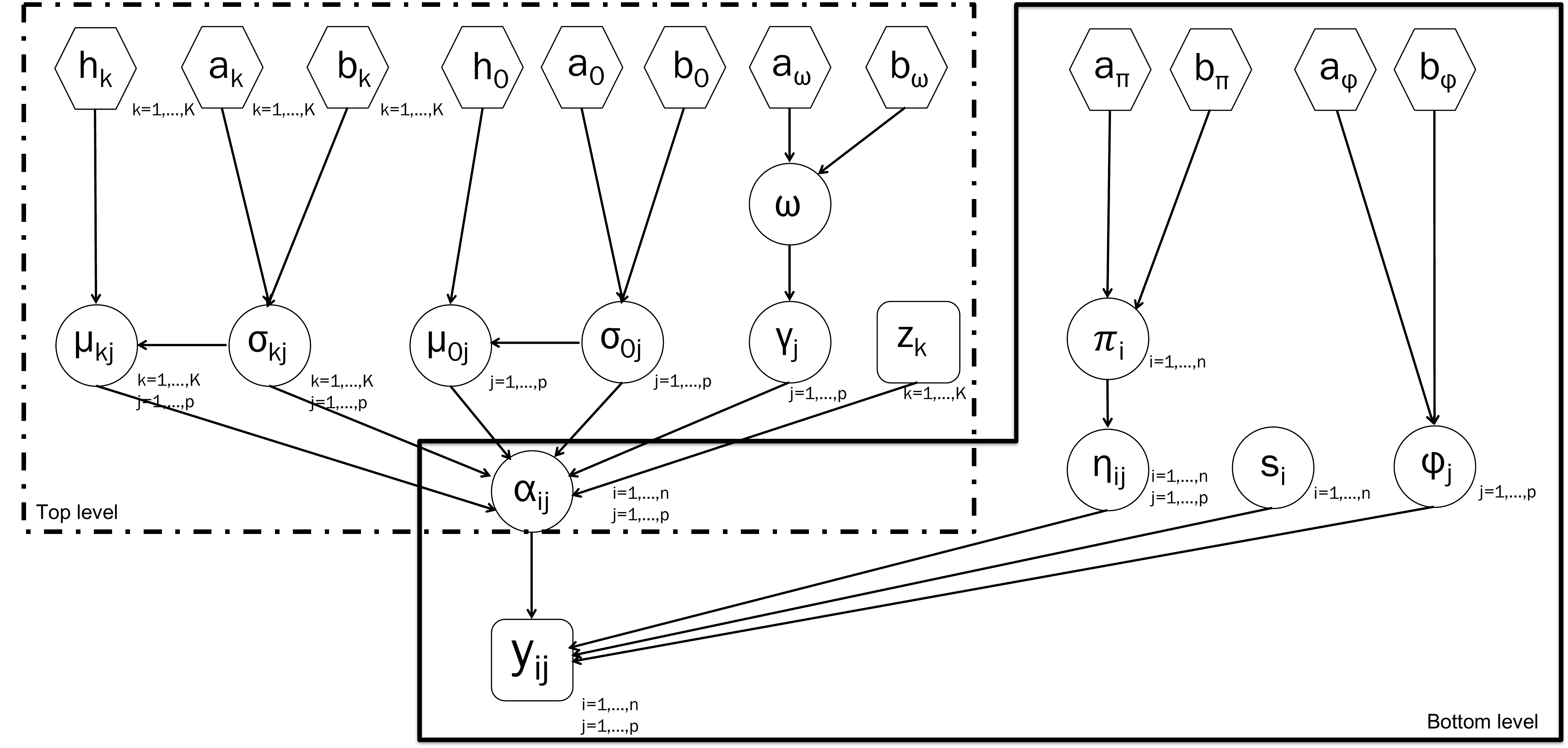}}
	\end{minipage}
	\caption{A graphical representation of the proposed bi-level Bayesian framework for microbial differential abundance analysis, with the bottom level (within the solid border) of (a) Dirichlet-multinomial (DM) model, and (b) zero-inflated negative binomial (ZINB) model. Each node in a circle/hexagon/square refers to a model parameter/a fixed hyperparameter/observable data. The link between two nodes represents a direct probabilistic dependence. Note that both (a) and (b) share the same top level (within the dashed border).}
	\label{f1}
\end{figure}

\section*{S2 \quad Details of the MCMC Algorithms}
We show the details of the MCMC algorithms of the proposed Bayesian framework, where phylogenetic structure is taken into account. For a simple cases where the data are only available at genus or OTU level for 16S rRNA sequencing data, or at species level for metagenomic shotgun sequencing data, please ignore the superscript $(2),\ldots,(l)$.

\subsection*{S2.1 \quad Bottom level}
\subsubsection*{S2.2.2 \quad Dirichlet-multinomial (DM) model}
We start by writing the likelihood for each sample $i,i=1,\ldots,n$, where the microbiome abundance is summarized at the bottom-most taxonomic levels, i.e. $l=1$,
\[f_\text{DM}(\bm{y}_{i\cdot}^{(1)}|\bm{\alpha}_{i\cdot}^{(1)})=\frac{\Gamma(Y_{i\cdot}+1)\Gamma(A_{i\cdot})}{\Gamma(Y_{i\cdot}+A_{i\cdot})}\prod_{j=1}^{p^{(1)}}\frac{\Gamma(y_{ij}^{(1)}+\alpha_{ij}^{(1)})}{\Gamma(y_{ij}^{(1)}+1)\Gamma(\alpha_{ij}^{(1)})}.\]
Note that $Y_{i\cdot}=\sum_{j=1}^{p^{(1)}}y_{ij}^{(1)}\cdots=\cdots\sum_{j=1}^{p^{(L)}}y_{ij}^{(L)}$ and $A_{i\cdot}=\sum_{j=1}^{p^{(1)}}a_{ij}^{(1)}\cdots=\cdots\sum_{j=1}^{p^{(L)}}a_{ij}^{(L)}$; that is, the total read counts and the total latent relative abundance should be unchanged, regardless of the choice of taxonomic levels. 

\subsubsection*{S2.1.2 \quad Zero-inflated negative binomial (ZINB) model}
We start by writing the likelihood for each sample $i,i=1,\ldots,n$, where the microbiome abundance is summarized at level $l$,
\[f_\text{ZINB}(\bm{y}_{i\cdot}^{(l)}|\bm{\alpha}_{i\cdot}^{(l)},\bm{\eta}_{i\cdot},\bm{\phi}^{(l)},s_i)=\prod_{j=1}^{p^{(l)}}f_\text{ZINB}({y}_{ij}^{(l)}|{\alpha}_{ij}^{(l)},{\eta}_{ij},\phi_j^{(l)},s_i),\]
where
\begin{align*}
&f_\text{ZINB}({y}_{ij}^{(l)}|{\alpha}_{ij}^{(l)},{\eta}_{ij},\phi_j^{(l)},s_i)\\
=&\text{I}(y_{ij}^{(1)}=0)^{\eta_{ij}}\left(\frac{\Gamma(y_{ij}^{(l)}+\phi_j^{(l)})}{y_{ij}^{(l)}!\Gamma(\phi_j^{(l)})}\left(\frac{\phi_j^{(l)}}{s_i\alpha_{ij}^{(1)}+\phi_j^{(l)}}\right)^{\phi_j^{(l)}}\left(\frac{s_i\alpha_{ij}^{(l)}}{s_i\alpha_{ij}^{(l)}+\phi_j^{(l)}}\right)^{y_{ij}^{(l)}}\right)^{1-\eta_{ij}}.
\end{align*}

{\bf Update of zero-inflation indicator $\eta_{ij}$:} We update each $\eta_{ij},i=1,\ldots,n,j=1,\ldots,p^{(1)}$ that corresponds to $y_{ij}^{(1)}=0$ by sampling from the normalized version of the following conditional:
\[p(\eta_{ij}|\cdot)\propto f_\text{ZINB}({y}_{ij}^{(1)}|{\alpha}_{ij}^{(1)},{\eta}_{ij},\phi_j,s_i)\cdot\text{Bern}(\eta_{ij};\pi_i).\]
After the Metropolis-Hasting steps for all $\eta_{ij}$, we use a Gibbs sampler to update each $\pi_i,i=1,\ldots,n$:
\[\pi_i|\cdot\sim \text{Be}(a_\pi+\sum_{j=1}^{p^{(1)}}\eta_{ij},b_\pi+p^{(1)}-\sum_{j=1}^{p^{(1)}}\eta_{ij}).\]

{\bf Update of dispersion parameter $\phi_{j}^{(l)}$:} We update each ${\phi}_{j}^{(l)},j=1,\ldots,p^{(l)},l=1,\ldots,L$ by using a random walk Metropolis-Hastings algorithm. We first propose a new ${\phi_{j}^{(l)}}^*$ from $\text{Ga}({\phi_{j}^{(l)}}^2/\tau_\phi,\phi_{j}^{(l)}/\tau_\phi)$ and then accept the proposed value ${\phi_{j}^{(l)}}^*$ with probability $\min(1,m_\text{MH})$, where
\begin{align*}
m_\text{MH}=\frac{\prod_{i=1}^nf_\text{ZINB}({y}_{ij}^{(l)}|{\alpha}_{ij}^{(l)},{\eta}_{ij},\phi_j^{(l)},s_i)}{\prod_{i=1}^nf_\text{ZINB}({y}_{ij}^{(l)}|{\alpha}_{ij}^{(l)},{\eta}_{ij},\phi_j^{(l)},s_i)}\frac{\text{Ga}({\phi_{j}^{(l)}}^*;a_\phi,b_\phi)}{\text{Ga}({\phi_{j}^{(l)}};a_\phi,b_\phi)}\frac{J({\phi_{j}^{(l)}};{\phi_{j}^{(l)}}^*)}{J({{\phi_{j}^{(l)}}^*;\phi_{j}^{(l)}})}.
\end{align*}
Here we use $J(\cdot|\cdot)$ to denote the proposal probability distribution for the selected move. Note that the last term, which is the proposal density ratio, can be canceled out for this random walk Metropolis update. 

{\bf Update of size factor $s_i$:} We can rewrite Equation (4) in the manuscript, i.e.
\[\log{s}_i \sim \sum_{m=1}^M\psi_m \left[t_m\, \text{N}(\nu_m,\sigma_s^2)+(1-t_m)\, \text{N}\left(-\frac{t_m\nu_m}{1-t_m},\sigma_s^2\right)\right]\]
by introducing latent auxiliary variables to specify how each sample (in terms of $\log s_i$) is assigned to any of the inner and outer mixture components. More specifically, we can introduce an $n\times 1$ vector of assignment indicators $\bm{g}$, with $g_i=m$ indicating that $\log{s}_i$ is a sample from the $m$-th component of the outer mixture. The weights $\psi_m$ determine the probability of each value $g_i=m$, with $m=1, \ldots,M$. Similarly, we can consider an $n\times 1$ vector $\bm{\epsilon}$ of binary elements $\epsilon_i$, where $\epsilon_i=1$ indicates that, given $g_i=m$, $\log{s}_i$ is drawn from the first component of the inner mixture, i.e. $\text{N}(\nu_m,\sigma_s^2)$ with probability $t_m$, and $\epsilon_i=0$ indicates that $\log{s}_i$ is drawn from the second component of the inner mixture, i.e. $\text{N}\left(-\frac{t_m\nu_m}{1-t_m},\sigma_s^2\right)$, with probability $1-t_m$. Thus, the Dirichlet process prior (DPP) model can be rewritten as
\begin{align*}
\log{s}_i|g_i,\epsilon_i,\bm{t},\bm{\nu} \sim \text{N}\left(\epsilon_i\nu_{g_i}+(1-\epsilon_i)\frac{-t_{g_i}\nu_{g_i}}{1-t_{g_i}},\sigma_s^2\right),
\end{align*}
where $\bm{t}$ and $\bm{\nu}$ denote the collections of $t_m$ and $\nu_m$, respectively. Therefore, the update of the size factor $s_i,i=1,\ldots,n$ can proceed by using a random walk Metropolis-Hastings algorithm. We propose a new $\log{s}_i^*$ from $\text{N}(\log{s}_i,\tau_s^2)$ and accept it with probability $\min(1,\text{m}_\text{MH})$, where
\[ \text{m}_\text{MH}=\frac{\prod_{j=1}^{p^{(1)}}f_\text{ZINB}({y}_{ij}^{(1)}|{\alpha}_{ij}^{(1)},{\eta}_{ij},\phi_j^{(1)},s_i^*)}{\prod_{j=1}^{p^{(1)}}f_\text{ZINB}({y}_{ij}^{(1)}|{\alpha}_{ij}^{(1)},{\eta}_{ij},\phi_j^{(1)},s_i)}\frac{\text{N}(\log{s}_i^*;\epsilon_i\nu_{g_i}+(1-\epsilon_i)\frac{-t_{g_i}\nu_{g_i}}{1-t_{g_i}},\sigma_s^2)}{\text{N}(\log{s}_i;\epsilon_i\nu_{g_i}+(1-\epsilon_i)\frac{-t_{g_i}\nu_{g_i}}{1-t_{g_i}},\sigma_s^2)}\frac{J(\log{s}_i;\log{s}_i^*)}{J(\log{s}_i^*;\log{s}_i)}.\]
Note that the last term, which is the proposal density ratio, equals $1$ for this random walk Metropolis update. Since $\bm{g}$, $\bm{\epsilon}$, $\bm{t}$, and $\bm{\nu}$ have conjugate full conditionals, we use Gibbs samplers to update them one after another:
\begin{itemize}
	\item Gibbs sampler for updating $g_i,i=1,\ldots,n$, by sampling from the normalized version of the following conditional:
	\[
	p(g_i=m|\cdot)\propto \psi_m\text{N}\left(\log s_i;\epsilon_i\nu_{m}+(1-\epsilon_i)\frac{-t_{m}\nu_{m}}{1-t_{m}},\sigma_s^2\right).
	\]
	\item Gibbs sampler for updating $\epsilon_i,i=1,\ldots,n$, by sampling from the normalized version of the following conditional:
	\[
	p(\epsilon_i|\cdot)\propto\begin{cases}{\begin{array}{ll} (1-t_m)\text{N}\left(\log s_i;-\frac{t_m\nu_m}{1-t_m},\sigma_s^2\right) & \textit{ if } \epsilon_i=0\\ t_m\text{N}\left(\log s_i;\nu_m,\sigma_s^2\right) & \textit{ if } \epsilon_i=1\end{array}}\end{cases}.
	\]
	\item Gibbs sampler for updating $t_m,m=1,\ldots,M$:
	\[
	t_m|\cdot\sim \text{Be}(a_t+\sum_{i=1}^n\text{I}(g_i=m)\text{I}(\epsilon_i=1),b_t+\sum_{i=1}^n\text{I}(g_i=m)\text{I}(\epsilon_i=0)).
	\]
	\item Gibbs sampler for updating $\nu_m,m=1,\ldots,M$:
	\[
	\nu_m|\cdot\sim\text{N}\left(\frac{c_m/\sigma_s^2}{e_m/\sigma_s^2+1/\tau_\nu^2},\frac{1}{e_m/\sigma_s^2+1/\tau_\nu^2}\right),
	\]
	where $c_m=\sum_{\{i:g_i=m,\epsilon_i=1\}}\log s_i-\frac{t_m}{1-t_m}\sum_{\{i:g_i=m,\epsilon_i=0\}}\log s_i$ and $e_m=\sum_{i=1}^n\text{I}(g_i=m)\text{I}(\epsilon_i=1)+\sum_{\{i:g_i=m,\epsilon_i=0\}}\left(\frac{t_m}{1-t_m}\right)^2$.
	\item Gibbs sampler for updating $\psi_m,m=1,\ldots,M$ by stick-breaking process \citep{Ishwaran2001}:
	\begin{align*}
	\psi_1 &= v_1,\\
	\psi_2 &= (1-v_1)v_2,\\
	&~~\vdots\\
	\psi_M &= (1-v_1)\cdots(1-v_{M-1})v_M,
	\end{align*}
	where $v_m|\bm{\nu}\sim \text{Be}\left(a_m+\sum_{i=1}^n\text{I}(g_i=m),b_m+\sum_{i=1}^n\text{I}(g_i>m)\right)$.
\end{itemize}

\subsection*{S2.2 \quad Top level}
Both of the DM model and the ZINB model share the same process to update the latent relative abundance matrix at the bottom-most taxonomic level, i.e. $\bm{A}^{(1)}$, and to select the discriminatory taxa at different levels, i.e. $\bm{\gamma}^{(1)},\ldots,\bm{\gamma}^{(L)}$. For the sake of convenience, we copy Equation (6) in the main text here,
\begin{align*}
&p(\bm{\alpha}_{\cdot j}^{(l)}|\gamma_j^{(l)})=(2\pi)^{-\frac{n}{2}}\times\\
&\begin{cases}
{\begin{array}{ll}
	\prod_{k=1}^K(n_kh_k+1)^{-\frac{1}{2}}\frac{\Gamma\left(a_k+\frac{n_k}{2}\right)}{\Gamma(a_k)}\frac{b_k^{a_k}}{\left\{b_k+\frac{1}{2}\left[\sum_{\{i:z_i=k\}}\log{\alpha_{ij}^{(l)}}^2-\frac{\left(\sum_{\{i:z_i=k\}}\log\alpha_{ij}^{(l)}\right)^2}{n_k+\frac{1}{h_k}}\right]\right\}^{a_k+\frac{n_k}{2}}} & \text{ if }  \gamma_j^{(l)}=1\\
	(nh_0+1)^{-\frac{1}{2}}\frac{\Gamma\left(a_0+\frac{n}{2}\right)}{\Gamma(a_0)}\frac{b_0^{a_0}}{\left\{b_0+\frac{1}{2}\left[\sum_{i=1}^n\log{\alpha_{ij}^{(l)}}^2-\frac{\left(\sum_{i=1}^n\log\alpha_{ij}^{(l)}\right)^2}{n+\frac{1}{h_0}}\right]\right\}^{a_0+\frac{n}{2}}} & \text{ if } \gamma_j^{(l)}=0\\
	\end{array}}.
\end{cases}
\end{align*} 

{\bf Update of relative abundance at the bottom-most level ${a}_{ij}^{(1)}$}: We update each ${\alpha}_{ij}^{(1)},i=1,\ldots,n,j=1,\ldots,p^{(1)}$ by using a Metropolis-Hastings random walk algorithm. We first propose a new ${{\alpha}_{ij}^{(1)}}^*$ from $\text{N}({\alpha}_{ij}^{(1)},\tau_\alpha^2)$, and then accept the proposed value with probability $\min(1,\text{m}_\text{MH})$, where
\[\text{m}_\text{MH}=\frac{f_\mathcal{M}(\bm{y}_{i\cdot}^{(1)}|{\bm{\alpha}_{i\cdot}^{(1)}}^*,\cdot)}{f_\mathcal{M}(\bm{y}_{i\cdot}^{(1)}|\bm{\alpha}_{i\cdot}^{(1)},\cdot)}\frac{p\left({\bm{\alpha}_{\cdot j}^{(1)}}^*|\gamma_j^{(1)}\right)}{p\left(\bm{\alpha}_{\cdot j}^{(1)}|\gamma_j^{(1)}\right)}\frac{J\left({\alpha}_{ij}^{(1)};{{\alpha}_{ij}^{(1)}}^*\right)}{J\left({{\alpha}_{ij}^{(1)}}^*;{\alpha}_{ij}^{(1)}\right)}.\]
Here we use $\mathcal{M}$ to denote the bottom-level model, which should be chosen from $\{\text{DM},\text{ZINB}\}$. Note that the last term, which is the proposal density ratio, equals $1$ for this random walk Metropolis update. 

{\bf Update of differentially abundant taxon indicator $\gamma_j^{(l)}$}: We update each $\gamma_j^{(l)},j=1,\ldots,p^{(l)},l=1,\ldots,L$ via an {\it add-delete} algorithm. In this approach, a new candidate vector, say ${\bm{\gamma}^{(l)}}^*$, is generated by randomly choosing an element within $\bm{\gamma}^{(l)}$, say $j$, and changing its value to $1-\gamma_j^{(l)}$. Then, this proposed move is accepted with probability $\text{min}(1,\text{m}_\text{MH})$, where the Hastings ratio is 
\[\text{m}_\text{MH}=\frac{p\left(\bm{\alpha}_{\cdot j}^{(l)}|{\gamma_j^{(l)}}^*\right)}{p\left(\bm{\alpha}_{\cdot j}^{(l)}|\gamma_j^{(l)}\right)}\frac{p\left({\gamma_j^{(l)}}^*|\cdot\right)}{p\left(\gamma_j^{(l)}|\cdot\right)} \frac{J\left(\bm{\gamma}^{(l)}|{\bm{\gamma}^{(l)}}^*\right)}{J\left({\bm{\gamma}^{(l)}}^*|\bm{\gamma}^{(l)}\right)}.\]	
Note that the proposal density ratio equals $1$. Here, we have two choices of $p\left(\gamma_j^{(l)}|\cdot\right)$: either independent Bernoulli prior or Markov random field prior (see Equation (7) in the manuscript). We should also notice that the feature selection and the abundance estimation are determined simultaneously in the MCMC algorithm. Therefore, to improve mixing, it is necessary to allow the selection to stabilize for any visited configurations of $\bm{A}^{(1)}$ and its induced $\bm{A}^{(l)}$'s. We suggest repeating the above Metropolis step multiple times within each iteration. In the simulations conducted for this paper, no improvement in the MCMC performance was noticed after repeating the step above $20$ times.

{\bf Update of relative abundance at upper levels ${a}_{ij}^{(l)},l\ge 2$}: For the DM model, the aggregation property can be used to derive the relative abundance at upper levels sequentially just from the one at the bottom level via $\alpha_{ij}^{(l)}=\sum_{\{j':g_{jj'}=1\}} \alpha_{ij'}^{(l-1)}$. For the ZINB model, the aggregation property does not hold. We assume that the size factor estimation should be irrelevant to the choices of microbiome count data at different taxonomic levels. Therefore, we update each ${\alpha}_{ij}^{(l)},i=1,\ldots,n,j=1,\ldots,p^{(l)},l=2,\ldots,L$ by using a Metropolis-Hastings random walk algorithm conditional on the size factors estimated by $\bm{Y}^{(1)}$. We first propose a new ${{\alpha}_{ij}^{(l)}}^*$ from $\text{N}({\alpha}_{ij}^{(l)},\tau_\alpha^2)$, and then accept the proposed value with probability $\min(1,\text{m}_\text{MH})$, where
\[\text{m}_\text{MH}=\frac{f_\text{ZINB}({y}_{ij}^{(l)}|{{\alpha}_{ij}^{(l)}}^*,{\eta}_{ij},\phi_j^{(l)},s_i)}{f_\text{ZINB}({y}_{ij}^{(l)}|{\alpha}_{ij}^{(l)},{\eta}_{ij},\phi_j^{(l)},s_i)}\frac{p\left({\bm{\alpha}_{\cdot j}^{(l)}}^*|\gamma_j^{(l)}\right)}{p\left(\bm{\alpha}_{\cdot j}^{(l)}|\gamma_j^{(l)}\right)}\frac{J\left({\alpha}_{ij}^{(l)};{{\alpha}_{ij}^{(l)}}^*\right)}{J\left({{\alpha}_{ij}^{(l)}}^*;{\alpha}_{ij}^{(l)}\right)}.\]
Note that the last term, which is the proposal density ratio, equals $1$ for this random walk Metropolis update.

\section*{S3 \quad Simulation}\label{S_simulation}
We use both simulated and synthetic data to assess the performance of the Bayesian framework embedded with the bottom-level model of DM and ZINB. We demonstrate the advantage of our models against alternative approaches. We also investigate how the prior choices affect the posterior inference.

\subsection*{S3.1 \quad Generative models}\label{generative}
Let $\bm{Y}_{n\times p}$ denote the simulated count table, where the number of features $p=1,000$, and the sample size $n=24$ or $108$. We do not consider the phylogenetic structure among the $p$ features in all simulation settings. We set the number of truly discriminatory taxonomic features $p_\gamma=50$ among $K=2$ or $3$ groups, helping us test the ability of our method to discover relevant features in the presence of a good amount of noise. The hierarchical formulations of the generative models are presented in Table \ref{simu.1}.

\begin{sidewaystable}
	%\begin{table}[h]
	\centering
	\tiny
	%\caption{The list of multivariate count generating process candidates $\mathcal{M}$}
	\begin{tabular}{|p{6cm}p{6cm}|p{6cm}|}
		\hline
		\multicolumn{2}{|c|}{Simulated data} &  \multicolumn{1}{c|}{Synthetic data}\\
		\hline
		{{The proposed DM Model}} & 
		{{The proposed ZINB Model}} &
		{The model proposed by {\cite{weiss2017normalization}}:} 
		\\\hline
		{{Bottom-level }:}
		{\begin{alignat*}{2}
			\text{For }& i=1,\ldots,n\\
			\bm{y}_i &\sim \text{Multi}(N_i, \psi_{i \cdot})\\
			&N_i \sim \text{U}(5,000, 10,000)\\
			&\psi_{i \cdot} \sim \text{Dir}(\bm{\alpha}_{i\cdot})
			\end{alignat*}}
		& 
		{\begin{alignat*}{2}
			\text{For }& i=1,\ldots,n,j=1,\ldots,p\\
			y_{ij} &\sim 0.5\text{I}(y_{ij}=0)+0.5\text{NB}(s_i\alpha_{ij}, \phi_{i})\\
			&s_i \sim \text{U}(0.5, 4)\\
			&\phi_{j} \sim \text{Exp}(1/10)\\
			\end{alignat*}} 
		&
		{{Count generative model}:}
		{\begin{alignat*}{2}
			\text{For }& i=1,\ldots,n\\
			\bm{y}_i & \sim \text{Multi}(10,000, \bm{\phi}_{i\cdot})\\ 
			& \bm{\psi}_{i\cdot}=\text{I}\left(1\le i\le \frac{n}{2} \right)\frac{\bm{P}}{\sum_{j=1}^pP_j}+\\
			& \quad\quad\quad\text{I}\left(\frac{n}{2}<i\le n\right)\frac{\bm{Q}}{\sum_{j=1}^pQ_j}
			\end{alignat*}}
		%	\vspace*{-100pt}
		\\
		\vspace{-27pt}
		\begin{multicols}{2}
			{{Top-level}:}
			{\begin{alignat*}{2}
				\log \alpha_{ij} &\sim
				\begin{cases}
				{\begin{array}{l}
					\text{I}\left(1\le i\le\frac{n}{2}\right)\text{N}\left(d_{1j},\sigma^2\right)+\text{I}\left(\frac{n}{2}< i\le n\right)\text{N}\left(d_{2j},\sigma^2\right),\\
					\quad\quad\text{ if }  \gamma_j=1 \text{ and } K=2\\
					\text{I}\left(1\le i\le\frac{n}{3}\right)\text{N}\left(d_{1j},\sigma^2\right)+\text{I}\left(\frac{n}{3}< i\le \frac{2n}{3}\right)\text{N}\left(d_{2j},\sigma^2\right)+I\left(\frac{2n}{3}< i\le n\right)\text{N}\left(d_{3j},\sigma^2\right),\\
					\quad\quad\text{ if }  \gamma_j=1 \text{ and } K=3\\
					\text{N}(d_{0j},\sigma^2/100),\\
					\quad\quad\text{ if } \gamma_j=0\\
					\end{array}}
				\end{cases} \\
				& d_{0j} \sim \text{U}(0,4)\\
				& \text{sort}(d_{1j},\ldots,d_{Kj}) =
				\begin{cases}
				{\begin{array}{l}
					(1-\sigma/2, 1+\sigma/2),\\
					\quad\quad\text{ if } K=2\\
					(1-\sigma, 1,  1+\sigma),\\\
					\quad\quad\text{ if } K=3\\
					\end{array}}
				\end{cases} \\
				\end{alignat*}}
		\end{multicols}
		&
		&
		{{Abundance generative model}:}
		{ \begin{alignat*}{2}
			P_j &= \begin{cases}
			\exp(\sigma) O_j & \text{ for } 1\le j\le p_\gamma/2\\ 
			O_j& \text{ otherwise } 
			\end{cases}\\
			Q_j &=\begin{cases}
			\exp(\sigma) O_j & \text{ for } p_\gamma/2<j\le p_\gamma\\ 
			O_j & \text{ otherwise }
			\end{cases}\\
			&\bm{O}=(O_1,\ldots,O_{p_\gamma/2},O_1,\ldots,O_{p_\gamma/2},O_{p_\gamma+1}\ldots,O_{p})^T,\\
			&\text{ which is summarized from real data}
			\end{alignat*}}
		\\\hline
	\end{tabular}
	\caption{Hierarchical formulations of the data generative models used in the simulation study. Note that the sample size $n\in\{24,108\}$, the number of features $p=1,000$ with $50$ discriminating among $K$ groups ($K\in\{2,3\}$ for simulated data and $K=2$ for synthetic data), and the effect size $\sigma\in\{1,2\}$.}
	\label{simu.1}
	%\end{table}
\end{sidewaystable}

\subsubsection*{S3.1.1 \quad  Generating simulated data}\label{generative_simulated}
We generated the simulated datasets that favor the proposed bi-level frameworks. For the latent relative abundance $\alpha_{ij}$ of a discriminating feature, we drew its logarithmic value from a two-component Gaussian mixture distribution,
\begin{align*}
\log{\alpha_{ij}}|\gamma_j=1\quad\sim\quad \text{I}\left(1\le i\le\frac{n}{2}\right)\text{N}\left(d_{1j},\sigma_\text{within}^2\right)+\text{I}\left(\frac{n}{2}< i\le n\right)\text{N}\left(d_{2j},\sigma_\text{within}^2\right)
\end{align*}
if $K=2$, or a three-component Gaussian mixture distribution,
\begin{align*}
\log{\alpha_{ij}}|\gamma_j=1\quad\sim\quad &\text{I}\left(1\le i\le\frac{n}{3}\right)\text{N}\left(d_{1j},\sigma_\text{within}^2\right)+\text{I}\left(\frac{n}{3}< i\le \frac{2n}{3}\right)\text{N}\left(d_{2j},\sigma_\text{within}^2\right)\\
&+\text{I}\left(\frac{2n}{3}< i\le n\right)\text{N}\left(d_{3j},\sigma_\text{within}^2\right)
\end{align*}
if $K=3$. Each permutation of $\{d_{1j},\ldots,d_{Kj}\}$ follows an arithmetic progression with unit mean and difference $\sigma$; that is, $\{1-\sigma/2,1+\sigma/2\}$ if $K=2$, and $\{1-\sigma, 1, 1+\sigma\}$ if $K=3$. For the scenario of $K=2$, $\sigma$ can be interpreted as the between-group standard deviation or the effect size in the logarithmic scale. We considered two scenarios of $\sigma=1$ or $2$, and set the within-group standard deviation $\sigma_\text{within}=\sigma/10$. For a non-discriminating feature, we generated its logarithmic value from a normal distribution with zero mean and variance $4$, i.e. $\log\alpha_{ij}|\gamma_j=0\sim\text{N}(0,4)$. For the bottom-level of the DM model, we first sampled the underlying fractional abundances for sample $i$ from a Dirichlet distribution with parameters $\bm{\alpha}_{i \cdot}$, i.e. $\bm{\psi}_{i \cdot}\sim\text{Dir}(\bm{\alpha}_{i \cdot})$. Then, their corresponding observed counts $\bm{y}_{i \cdot}$ were drawn from a multinomial distribution, i.e. $\text{Multi}(N_i,\bm{\psi}_{i \cdot})$, where the total counts $N_i$ was randomly selected from a discrete uniform distribution $\text{U}(50,000,10,000)$. As for the ZINB model, we sampled the size factors $s_i$ from a uniform distribution $\text{U}(0.5,4)$, and the dispersion parameters $\phi_j$ from an exponential distribution with mean $10$, i.e. $\text{Exp}(1/10)$. Next, each observed count $y_{ij}$ was generated from $\text{NB}(s_i\alpha_{ij},\phi_j)$. Lastly, we randomly selected half of the counts and forced their values to zero in order to mimic the excess zeros seen in the real data. Combined with the two bottom-level kernels ($\{\text{DM},\text{ZINB}\}$), the two choices of the sample size ($n\in\{24,108\}$), the number of groups ($K\in\{2,3\}$) and the $\log$ effect size ($\sigma\in\{1,2\}$), there were $2^4=16$ scenarios in total. For each of the scenarios, we independently repeated the above steps to generate $50$ datasets.

\subsubsection*{S3.1.2 \quad  Generating synthetic data}\label{generative_synthetic}
To evaluate the performance of the proposed methods on the count data that are different from the model assumptions, we also generated synthetic datasets based on multinomial models that characterize a real taxa abundance distribution. A brief description of the data-generating scheme is given below, while detailed information can be found in the supplement of \cite{weiss2017normalization}. Let $\bm{O}=(O_1,\ldots,O_{p_\gamma/2},O_{p_\gamma/2+1},\ldots,O_{p_\gamma},O_{p_\gamma+1}\ldots,O_{p})^T$ be a count vector, where $(O_1,\ldots,O_{p_\gamma/2})=(O_{p_\gamma/2+1},\ldots,O_{p_\gamma})$, and each $O_j,p_\gamma/2<j\le p$ is the sum of OTU counts for one randomly selected taxon (without replacement) from all the skin or feces samples in a real microbiome study \citep{caporaso2011global}. We define two $p$-by-$1$ vectors, $\bm{P}$ and $\bm{Q}$, as
\begin{align*}
P_j = \begin{cases}
\exp(\sigma) O_j & \text{ for } 1\le j\le p_\gamma/2\\ 
O_j& \text{ otherwise } 
\end{cases}, ~\text{and}~
Q_j =\begin{cases}
\exp(\sigma) O_j & \text{ for } p_\gamma/2<j\le p_\gamma\\ 
O_j & \text{ otherwise }
\end{cases},
\end{align*}
where $\sigma$ represents the $\log$ effect size. Note that $\sum_{j=1}^pP_j=\sum_{j=1}^pQ_j$. We further drew the observed counts $\bm{y}_{i\cdot}$ from a multinomial model $\text{Multi}(N_i,\bm{\psi}_{i\cdot})$, where $N_i=10,000$ and $\bm{\psi}_{i\cdot}  =  \text{I}\left(1\le i\le \frac{n}{2} \right)\frac{\bm{P}}{\sum_{j=1}^pP_j}+\text{I}\left(\frac{n}{2}<i\le n\right)\frac{\bm{Q}}{\sum_{j=1}^pQ_j}$.
This would yield the first $p_{\gamma}$ taxa to be truly discriminating between the two equally sized groups. Finally, we permuted the columns of the data matrix, $\bm{Y}$, to disperse the taxa. Combined with the two types of samples ($\{\text{Skin},\text{Feces}\}$), and the two choices of the sample size ($n\in\{24,108\}$) and the $\log$ effect size ($\sigma\in\{1,2\}$), there were $2^3=8$ scenarios in total. For each of the scenarios, we repeated the steps above to generate $50$ independent datasets.

\subsection*{S3.2 \quad  Algorithm settings}\label{algo_set}
For prior specification in the top level of the proposed Bayesian framework, we used the following default settings. We set the hyperparameters that control the selection of discriminatory features, $\omega\sim\text{Be}(a_\omega=0.2,b_\omega=1.8)$, resulting in the proportion of taxa expected \textit{a priori} to discriminate among the $K$ groups to be $a_\omega/(a_\omega+b_\omega)=10\%$. As for the inverse-gamma priors on the variance components $\sigma_{0j}^2$ and $\sigma_{kj}^2$, we set the shape parameters $a_0=a_1=\ldots=a_k = 2$ and the scale parameters $b_0=b_1=\ldots=b_k=1$ to achieve a fairly flat distribution with an infinite variance. We further set the default values of $h_0$ and $h_k$ to $100$, as our sensitivity analysis in Section \ref{sa} shows the posterior inference on $\bm{\gamma}$ remained almost the same when those values were in the range of $10$ to $100$. As indicated by \cite{stingo2013integrative}, larger values of these hyperparameters would encourage the selection of only very large effects whereas smaller values would encourage the selection of smaller effects. For the bottom level of the ZINB model, we used the following weakly informative settings. The hyperparameters that control the percentage of extra zeros \textit{a priori} were set to $\pi\sim\text{Be}(a_\pi=1,b_\pi=1)$. As for the gamma prior on the dispersion parameters, i.e. $\phi_j\sim\text{Ga}(a_\phi,b_\phi)$, we set both $a_{\phi}$ and $b_{\phi}$ to small values such as $0.001$, which leads to a vague prior with expectation and variance equal to $1$ and $1,000$. For the Dirichlet priors on the size factors $s_i$, we followed \cite{Li2017} by specifying $M = n/2, ~\sigma_s = 1, ~\tau_{\eta} = 1, ~a_t = b_t = 1$, and $a_m = b_m = 1$. For each dataset, we ran a MCMC chain with $10,000$ iterations (first half as burn-in). The chain was initialed from a model with $5\%$ randomly chosen $\gamma_j$ set to $1$. Note that the DM model does not have any parameters needing to be specified in the bottom level. 

\subsection*{S3.3 \quad  Performance metrics}
To quantify the accuracy of identifying discriminatory features via the binary vector $\bm{\gamma}$, we used two widely used measures of the quality of binary classifications: 1) area under the curve (AUC) of the receiver operating characteristic (ROC); and 2) Matthews correlation coefficient (MCC) \citep{matthews1975comparison}. The former considers both true positive (TP) and false positive (FP) rates across various threshold settings, while the latter balances TP, FP, true negative (TN), and false negative (FN) counts even if the true zeros and ones in $\bm{\gamma}$ are of very different sizes. MCC is defined as \[\frac{(\text{TP}\times \text{TN} - \text{FP}\times \text{FN})}{\sqrt{(\text{TP+FP})(\text{TP+FN})(\text{TN+FP})(\text{TN+FN})}}.\] In differential analysis settings, the number of truly discriminatory features are usually assumed to be a small fraction to the total. Therefore, MCC is more appropriate to handle such an imbalanced scenario. Note that the AUC yields a value between $0$ to $1$ that is averaged by all possible thresholds that are used to select discriminatory taxa based on PPI, and the MCC value ranges from $-1$ to $1$ to pinpoint a specified threshold. The larger the index, the more accurate the inference.

\subsection*{S3.4 \quad  Results}\label{simu_res}
\begin{figure}[!h]
	\centering
	\includegraphics[width=.8\textwidth]{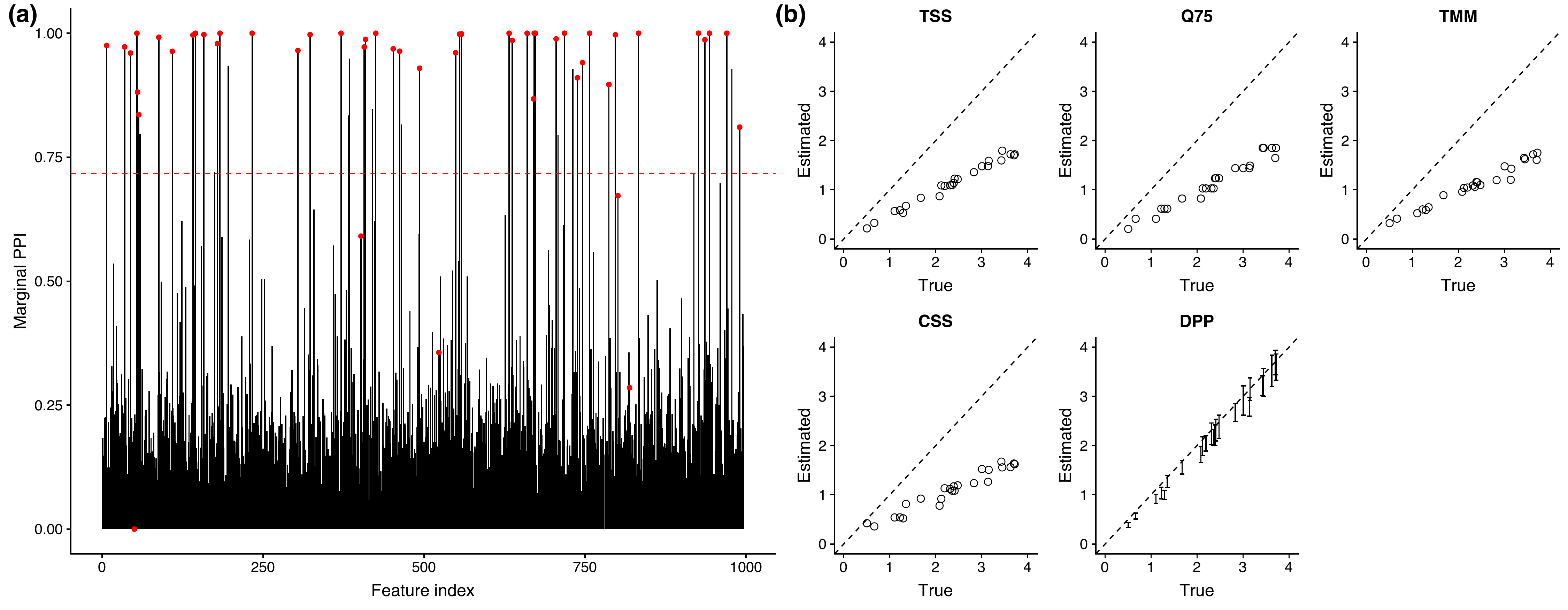} 
	\caption{Simulated data: (a) Marginal posterior probabilities of inclusion (PPI), $p(\gamma_j|\cdot)$, with the red dots indicating truly discriminatory features and the horizontal red dashed line indicating a threshold for a $5\%$ Bayesian FDR; (b) The scatter plots of the true and estimated size factors $s_i$'s obtained by different normalization methods. Note that RLE is not shown here because a large number of zeros in the data made the geometric means (the key component to calculating the size factors) of a few features inadmissible.}
	\label{sizefactor}
\end{figure}

We first describe posterior inference on the parameters of interest, $\bm{\gamma}$ and $\bm{s}$, on a single simulated dataset (bottom-level kernel$=$ZINB, $n=24$, $K=2$, $\sigma=2$). The results were obtained by fitting the proposed framework where the bottom level is a ZINB model with DPP as the normalization method, denoted by ZINB-DPP. As for the feature selection, Figure \ref{sizefactor}(a) shows the marginal PPI of each feature, $p(\gamma_j|\cdot)$. The red dots indicate the truly discriminatory features and the horizontal dashed line corresponds to a threshold that ensures an expected Bayesian FDR of $5\%$. This threshold resulted in a model that included $55$ features, $45$ of which were in the set of truly discriminatory features. As for the size factors $\bm{s}$, we plot the true values against the estimated ones by different normalization techniques in Figure \ref{sizefactor}(b). One advantage of the use of DPP is that it can output the uncertainty of our estimation on the size factors $\bm{s}$. It clearly shows that all of the true values were within the $95\%$ credible intervals derived by our method. Note that the true size factors were generated from $\text{U}(0.5, 4)$ instead of the mixture model that DPP assumes. In comparison, the alternative normalization techniques with constraint $\prod_{i=1}^ns_i=1$ yielded biased estimations.

To demonstrate the superiority of the proposed Bayesian models, particularly the ZINB model where the DPP is used to normalize the samples, we compared ours with other general approaches for microbial differential abundance analysis, all of which can be easily implemented in \texttt{R}. They are: 1) Analysis of variance (ANOVA); 2) Kruskal-Wallis test; 3) \texttt{edgeR} \citep{robinson2010edgeR}; 4) \texttt{DESeq2} \citep{love2014moderated}; and 5) \texttt{metagenomeSeq} \citep{Paulson2013}. The first two are parametric/nonparametric methods for testing whether samples originate from the same distribution, after converting each $\bm{y}_{i\cdot}$ into a compositional vector of proportions by dividing each count by the total number of reads $Y_{i\cdot}$. Note that the aim is to determine whether there is a significant difference among the abundance means/medians of multiple groups for each individual taxonomic feature. The third and fourth ones were developed for the analysis of RNA-Seq count data, but can be used to analyze microbiome data. \texttt{edgeR} implements an exact binomial test generalized for over-dispersed counts, while \texttt{DESeq2} employs a Wald test by adopting a generalized linear model based on an NB kernel. The last one, \texttt{metagenomeSeq}, assumes a zero inflated Gaussian model on the log-transformed counts, and performs a multiple groups test on moderated F-statistics. All these competitors produce $p$-values. In order to control for the FDR, i.e. the rate of type-I errors in these null hypothesis testings, we further adjusted their $p$-values using the BH method \citep{benjamini1995controlling}. As mentioned in Section \ref{generative}, we independently generated $50$ replicates for each of the $16$ simulated data scenarios, and each of the eight synthetic data scenarios. For each dataset, we ran the DM and ZINB-DPP models, and the five competitors, and computed their individual AUC and MCC. 
\begin{figure}[!h]
	\centering
	\includegraphics[width=.8\textwidth]{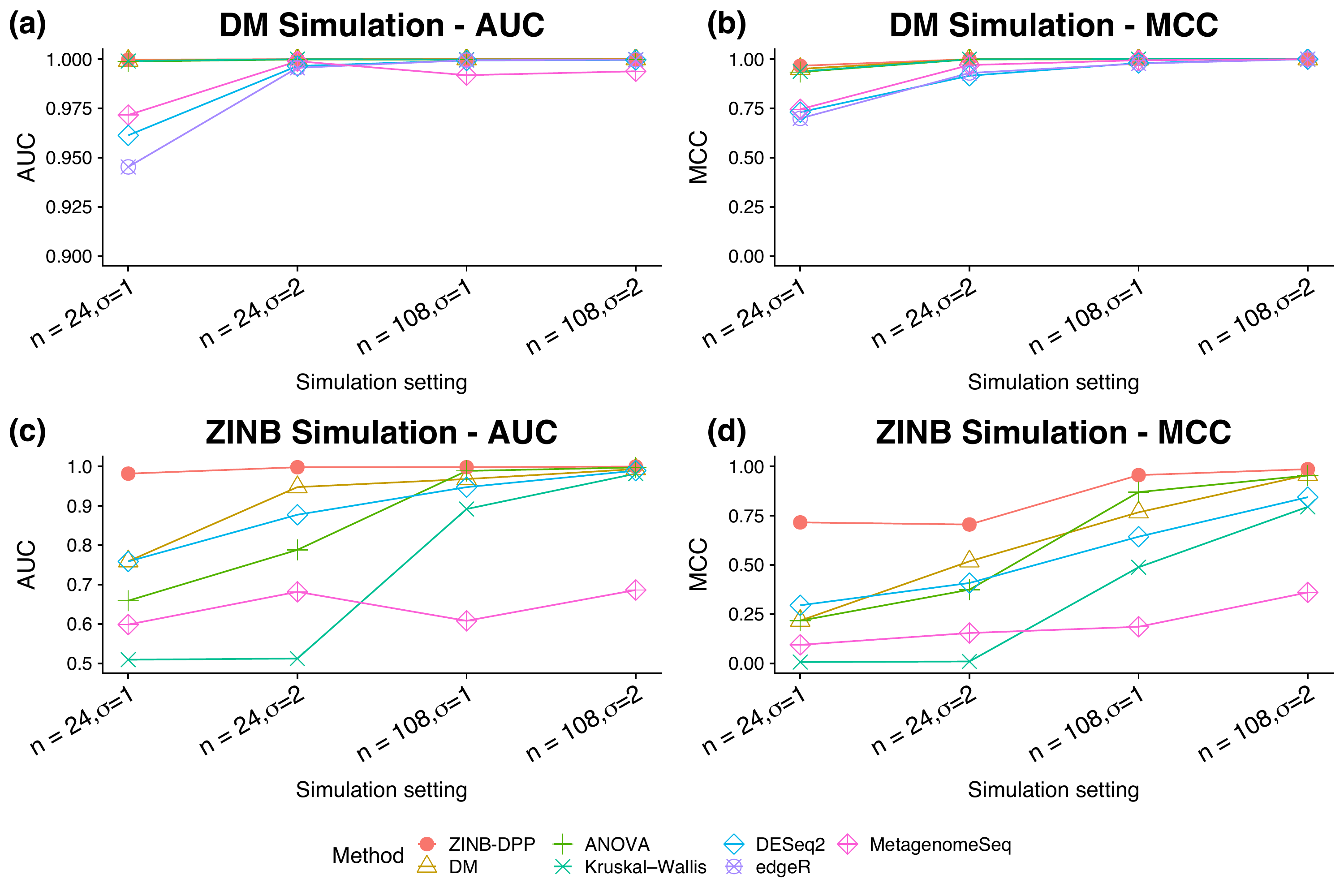} 
	\caption{Simulation study: Averaged AUC and MCC achieved by the proposed framework with DM and ZINB models, and the five competitors: ANOVA, Kruskal-Wallis, \texttt{edgeR}, \texttt{DESeq2}, and \texttt{metagenomeSeq}. A and B are plotted from the simulated data generated by the DM model. C and D are plotted from the simulated data generated by the ZINB model. }
	\label{simu_fig_1}
\end{figure}

The average AUC over $50$ simulated datasets under the same group number ($K=3$) and different sample sizes and effect sizes, $(n,\sigma)$, are displayed in Figure \ref{simu_fig_1}(a) and \ref{simu_fig_1}(c). It shows that all methods performed reasonably well for the data generated by the DM model when either the sample size or the effect size was fairly large. However, for a small sample size ($n=24$) and a small effect size $(\sigma=1)$, the performance of \texttt{edgeR}, \texttt{DESeq2}, and \texttt{metagenomeSeq} significantly dropped. For the data generated by the ZINB model, the results show that the ZINB-DPP model always achieved the highest AUC values. Decreasing either the sample size or the effect size would lead to greater disparity between ZINB-DPP and the others. Figure \ref{simu_fig_1}(b) and \ref{simu_fig_1}(d) show the comparison in terms of MCC. To make a fair comparison between the methods that output $p$-values and those that output probability measures such as PPI, we picked only the top $50$ significant features of each dataset from each method and computed their individual MCC. These two plots confirm the overall best performance of our proposed ZINB-DPP model. The related numerical results are summarized in Tables \ref{simuauc} and \ref{simumcc}, which show the model performance with respect to AUC and MCC on the simulated data generated from the DM and ZINB models. Note that ZINB-RLE and \texttt{edgeR} failed to produce any results on data generated by the ZINB model. This is because a large number of zeros is likely to make the geometric means (the key component to calculating the size factors) of a few features inadmissible. 

\begin{table}
	\centering
	\resizebox{\columnwidth}{!}{%
		\begin{tabular}{c|c|c|c|c|c|c|c|c|c}
			\hline
			\hline
			\multirow{4}{*}{\pbox{20cm}{\shortstack{Generative \\ Model}}}
			& \multirow{4}{*}{Methods} & \multicolumn{8}{c}{Simulation Setting} \\
			\cline{3-10} 
			& & \multicolumn{4}{c|}{$K =2$} & \multicolumn{4}{c}{$K =3$} \\
			\cline{3-10} 
			& & \multicolumn{2}{c}{$n =24$} & \multicolumn{2}{c|}{$n =108$} & 
			\multicolumn{2}{c}{$n =24$} & \multicolumn{2}{c}{$n =108$}  \\
			\cline{1-10} 
			& & $\sigma = 1$ & $\sigma = 2$ & $\sigma = 1$ & $\sigma = 2$ & $\sigma = 1$ & $\sigma = 2$ & $\sigma = 1$ & $\sigma = 2$ \\
			\cline{3-10} 
			\multirow{13}{*}{DM} & \multirow{2}{*}{ZINB-DPP} & \textbf{1.000} & \textbf{1.000} & \textbf{1.000} & \textbf{1.000} & \textbf{1.000}& \textbf{1.000} & \textbf{1.000} & \textbf{1.000} \\
			[-2ex]& & (0.0102) & (0.0000) & (0.0000) & (0.0000) & (0.0003) & (0.0000) & (0.0000) & (0.0000) \\ 
			\cline{2-10} 
			& \multirow{2}{*}{ZINB-TSS} & 0.977 & 1.000 & 1.000 & 1.000 & 0.999 & 1.000 & 1.000 & 1.000 \\ 
			[-2ex]& & (0.0111) & (0.0002) & (0.0002) & (0.0001) & (0.0006) & (0.0002) & (0.0001) & (0.0002)\\ 
			\cline{2-10} 
			& \multirow{2}{*}{ZINB-Q75} & 0.977 & 1.000 & 1.000 & 1.000 & 0.999 & \textbf{1.000} & \textbf{1.000} & 1.000 \\ 
			[-2ex]& & (0.0112) & (0.0001) & (0.0001) & (0.0001) & (0.0005) & (0.0000) & (0.0000) & (0.0001)\\ 
			\cline{2-10} 
			& \multirow{2}{*}{ZINB-RLE} & 0.978 & 1.000 & 1.000 & \textbf{1.000} & 1.000 & \textbf{1.000} & 1.000 & 1.000 \\ 
			[-2ex]& & (0.0100) & (0.0001) & (0.0001) & (0.0000) & (0.0004) & (0.0000) & (0.0001) & (0.0001)\\ 
			\cline{2-10} 
			& \multirow{2}{*}{ZINB-TMM} & 0.978 & 1.000 & 1.000 & \textbf{1.000} & 1.000 & \textbf{1.000} & 1.000 & 1.000 \\ 
			[-2ex]& & (0.0105) & (0.0001) & (0.0001) & (0.0000) & (0.0005) & (0.0000) & (0.0001) & (0.0001)\\ 
			\cline{2-10} 
			& \multirow{2}{*}{ZINB-CSS} & 0.976 & \textbf{1.000} & 1.000 & 1.000 & 0.999 & \textbf{1.000} & 1.000 & 1.000 \\ 
			[-2ex]& & (0.0115) & (0.0000) & (0.0000) & (0.0001) & (0.0008) & (0.0000) & (0.0001) & (0.0001)\\ 
			\cline{2-10} 
			& \multirow{2}{*}{DM} & 0.972 & 1.000 & 1.000 & 1.000 & 0.999 & 1.000 & 1.000 & 1.000 \\ 
			[-2ex]& & (0.0111) & (0.0002) & (0.0002) & (0.0001) & (0.0006) & (0.0002) & (0.0001) & (0.0002)\\ 
			\cline{2-10} 
			& \multirow{2}{*}{ANOVA} & 0.977 & 1.000 & 1.000 & 1.000 & 0.999 & 1.000 & 1.000 & 1.000 \\ 
			[-2ex]& & (0.0101) & (0.0003) & (0.0003) & (0.0004) & (0.0010)& (0.0004) & (0.0003) & (0.0004)\\ 
			\cline{2-10} 
			& \multirow{2}{*}{Kruskal–Wallis} & 0.978 & 1.000 & 1.000 & 1.000 & 0.999 & 1.000 & 1.000 & 1.000\\ 
			[-2ex]& & (0.0102) & (0.0002) & (0.0003) & (0.0004) & (0.0008) & (0.0001) & (0.0003) & (0.0003)\\ 
			\cline{2-10} 
			& \multirow{2}{*}{ \texttt{DESeq2}} & 0.981 & 1.000 & 1.000 & 0.999 & 0.961 & 0.997 & 0.999 & 1.000\\ 
			[-2ex]& & (0.0094) & (0.0005) & (0.0004) & (0.0004) & (0.0154) & (0.0033) & (0.0008) & (0.0004)\\ 
			\cline{2-10} 
			& \multirow{2}{*}{ \texttt{edgeR}} & 0.969 & 0.999 & 1.000 & 1.000 & 0.945 & 0.996 & 0.999 & 1.000\\ 
			[-2ex]& & (0.0130) & (0.0006) & (0.0005) & (0.0006) & (0.0237) & (0.0049) & (0.0007) & (0.0004)\\ 
			\cline{2-10} 
			& \multirow{2}{*}{ \texttt{metagenomeSeq}} & 0.957 & 0.998 & 0.895 & 0.918 & 0.972 & 0.999 & 0.992 & 0.994\\ 
			[-2ex]& & (0.0146) & (0.0017) & (0.0239) & (0.0199) & (0.0122) & (0.0015) & (0.0043) & (0.0040)\\ 
			\cline{1-10} 
			\multirow{13}{*}{ZINB} & \multirow{2}{*}{ZINB-DPP} & \textbf{0.907} & \textbf{0.990} & \textbf{0.994} & \textbf{0.998} & \textbf{0.982} & 0.998 & \textbf{0.998} & \textbf{1.000}\\
			[-2ex]& & (0.0203) & (0.0095) & (0.0051) & (0.0039) & (0.0085) & (0.0072) & (0.0026) & (0.0003)\\ 
			\cline{2-10} 
			& \multirow{2}{*}{ZINB-TSS} & 0.888 & 0.988 & 0.993 & 0.997 & 0.975 & \textbf{0.998} & 0.997 & 1.000 \\ 
			[-2ex]& & (0.0240) & (0.0097) & (0.0059) & (0.0047)& (0.0125) & (0.0065) & (0.0044) & (0.0005)\\ 
			\cline{2-10} 
			& \multirow{2}{*}{ZINB-Q75} & 0.888 & 0.988 & 0.991 & 0.997 & 0.975 & 0.998 & 0.997 & 1.000 \\ 
			[-2ex]& & (0.0225) & (0.0101) & (0.0070) & (0.0051) & (0.0106) & (0.0043) & (0.0043) & (0.0005)\\ 
			\cline{2-10} 
			& \multirow{2}{*}{ZINB-RLE} & NA & NA &NA  & NA  & NA  & NA  & NA  & NA  \\ 
			[-2ex]& & (-) & (-) & (-) & (-) & (-) & (-) & (-) & (-)\\ 
			\cline{2-10} 
			& \multirow{2}{*}{ZINB-TMM} & 0.887 & 0.988 & 0.992 & 0.997 & 0.974 & 0.998 & 0.997 & \textbf{1.000} \\ 
			[-2ex]& &(0.0247) & (0.0103) & (0.0064) & (0.0045) & (0.0116) & (0.0062) & (0.0040) & (0.0003)\\ 
			\cline{2-10} 
			& \multirow{2}{*}{ZINB-CSS} &0.881 & 0.988 & 0.991 & 0.997 & 0.973 & 0.998 & 0.997 & \textbf{1.000} \\ 
			[-2ex]& & (0.0245) & (0.0094) & (0.0073) & (0.0049) & (0.0128) & (0.0046) & (0.0047) & (0.0003)\\ 
			\cline{2-10} 
			& \multirow{2}{*}{DM} &0.659 & 0.856 & 0.929 & 0.990 & 0.759 & 0.947 & 0.968 & 0.993 \\ 
			[-2ex]& & (0.0378) & (0.0261) & (0.0173) & (0.0109) & (0.0499) & (0.0245) & (0.0175) & (0.0098)\\ 
			\cline{1-10} 
		\end{tabular}
	}
\end{table}

\begin{table}
	\centering
	\resizebox{\columnwidth}{!}{%
		\begin{tabular}{c|c|c|c|c|c|c|c|c|c}
			\hline
			\hline
			\multirow{4}{*}{\pbox{20cm}{\shortstack{Generative \\ Model}}}
			& \multirow{4}{*}{Methods} & \multicolumn{8}{c}{Simulation Setting} \\
			\cline{3-10} 
			& & \multicolumn{4}{c|}{$K =2$} & \multicolumn{4}{c}{$K =3$} \\
			\cline{3-10} 
			& & \multicolumn{2}{c}{$n =24$} & \multicolumn{2}{c|}{$n =108$} & 
			\multicolumn{2}{c}{$n =24$} & \multicolumn{2}{c}{$n =108$}  \\
			\cline{1-10} 
			& & $\sigma = 1$ & $\sigma = 2$ & $\sigma = 1$ & $\sigma = 2$ & $\sigma = 1$ & $\sigma = 2$ & $\sigma = 1$ & $\sigma = 2$ \\
			\cline{3-10} 
			\multirow{6}{*}{ZINB} & \multirow{2}{*}{ANOVA} & 0.714 & 0.908 & 0.972 & 0.995 & 0.659 & 0.788 & 0.989 & 0.997 \\ 
			[-2ex]& & (0.0601) & (0.0333) & (0.0126) & (0.0052) & (0.0689) & (0.0608) & (0.0066) & (0.0031)\\ 
			\cline{2-10} 
			& \multirow{2}{*}{Kruskal–Wallis} & 0.547 & 0.634 & 0.824 & 0.942 & 0.509 & 0.512 & 0.892 & 0.982\\ 
			[-2ex]& & (0.0528)& (0.0834) & (0.0331) & (0.0197) & (0.0211) & (0.0216) & (0.0299) & (0.0098)\\ 
			\cline{2-10} 
			& \multirow{2}{*}{ \texttt{DESeq2}} & 0.761 & 0.937 & 0.969 & 0.993 & 0.759 & 0.877 & 0.947 & 0.989\\ 
			[-2ex]& & (0.0369) & (0.0239) & (0.0151) & (0.0077) & (0.0429) & (0.0418) & (0.0255) & (0.0079)\\ 
			\cline{2-10} 
			& \multirow{2}{*}{ \texttt{edgeR}} & NA & NA &NA  & NA  & NA  & NA  & NA  & NA  \\ 
			[-2ex]& & (-) & (-) & (-) & (-) & (-) & (-) & (-) & (-)\\ 
			\cline{2-10} 
			& \multirow{2}{*}{ \texttt{metagenomeSeq}} & 0.609 & 0.766 & 0.750 & 0.933 & 0.599 & 0.682 & 0.608 & 0.686\\ 
			[-2ex]& & (0.0410) & (0.0443) & (0.0401) & (0.0252) & (0.0555) & (0.0697) & (0.0493) & (0.0535)\\ 
			\hline
			\hline 
		\end{tabular}
	}
	\caption{DM and ZINB simulation: area under the curve (AUC) given by all methods. In each cell, the top number is the averaged AUC over 50 independent datasets, and the bottom number in parentheses is the standard error. The result from the model that achieved best performance under the associated scenario (each column) is marked in bold.}
	\label{simuauc}
\end{table}

\begin{table}
	\centering
	\resizebox{\columnwidth}{!}{%
		\begin{tabular}{c|c|c|c|c|c|c|c|c|c}
			\hline
			\hline
			\multirow{4}{*}{\pbox{20cm}{\shortstack{Generative \\ Model}}}
			& \multirow{4}{*}{Methods} & \multicolumn{8}{c}{Simulation Setting} \\
			\cline{3-10} 
			& & \multicolumn{4}{c}{$K =2$} & \multicolumn{4}{c}{$K =3$} \\
			\cline{3-10} 
			& & \multicolumn{2}{c}{$n =24$} & \multicolumn{2}{c}{$n =108$} & 
			\multicolumn{2}{c}{$n =24$} & \multicolumn{2}{c}{$n =108$}  \\
			\cline{1-10} 
			& & $\sigma = 1$ & $\sigma = 2$ & $\sigma = 1$ & $\sigma = 2$ & $\sigma = 1$ & $\sigma = 2$ & $\sigma = 1$ & $\sigma = 2$ \\
			\cline{3-10} 
			\multirow{13}{*}{DM} & \multirow{2}{*}{ZINB-DPP} & \textbf{0.800} &  0.997 & 0.999 & 1.000 & \textbf{0.966} & \textbf{1.000} & \textbf{1.000}  & \textbf{1.000}  \\
			[-2ex]& & (0.0379) & (0.0074) & (0.0042) & (0.0000) & (0.0180) & (0.0000) & (0.0000) & (0.0000) \\ 
			\cline{2-10} 
			& \multirow{2}{*}{ZINB-TSS} & 0.779 & 0.996 & 0.999 & 1.000 & 0.954 & \textbf{1.000}  & 1.000 & \textbf{1.000}  \\ 
			[-2ex]& & (0.0474) & (0.0082) & (0.0042) & (0.0030) & (0.0260) & (0.0000) & (0.0030) & (0.0000)\\ 
			\cline{2-10} 
			& \multirow{2}{*}{ZINB-Q75} & 0.777 & 0.996 & 0.999 & 0.998 & 0.957 & 1.000 & 1.000 & 1.000 \\ 
			[-2ex]& & (0.0479) & (0.0085) & (0.0051) & (0.0058) & (0.0267) & (0.0000) & (0.0030) & (0.0030)\\ 
			\cline{2-10} 
			& \multirow{2}{*}{ZINB-RLE} & 0.782 & 0.995 & 1.000 & 1.000 & 0.960 & 1.000 & 0.999 & 1.000 \\ 
			[-2ex]& & (0.0495) & (0.0091) & (0.0030) & (0.0000) & (0.0220) & (0.0000) & (0.0042) & (0.0030)\\ 
			\cline{2-10} 
			& \multirow{2}{*}{ZINB-TMM} & 0.776 & 0.995 & 0.999 & 0.999 & 0.957 & \textbf{1.000}  & 1.000 & \textbf{1.000}  \\ 
			[-2ex]& & (0.0451) & (0.0091) & (0.0042) & (0.0042) & (0.0224) & (0.0000) & (0.0030) & (0.0000)\\ 
			\cline{2-10} 
			& \multirow{2}{*}{ZINB-CSS} & 0.774 & 0.995 & 1.000 & 0.999 & 0.955 & 1.000 & 0.999 & 1.000 \\ 
			[-2ex]& & (0.0494) & (0.0093) & (0.0030) & (0.0042) & (0.0266) & (0.0035) & (0.0042) & (0.0030)\\ 
			\cline{2-10} 
			& \multirow{2}{*}{DM} & 0.751 & 0.997 & 0.999 & 0.999 & 0.951 & 0.999 & 0.999 & 0.999 \\ 
			[-2ex]& & (0.0515) & (0.0074) & (0.0042) & (0.0051) & (0.0231) & (0.0037) & (0.0051) & (0.0042)\\ 
			\cline{2-10} 
			& \multirow{2}{*}{ANOVA} & 0.739 & 0.982 & \textbf{1.000}  & \textbf{1.000}  & 0.935 & 0.999 & \textbf{1.000}  & \textbf{1.000}  \\ 
			[-2ex]& & (0.0529) & (0.0153) & (0.0000) & (0.0000) & (0.0280) & (0.0052) & (0.0000) & (0.0000)\\ 
			\cline{2-10} 
			& \multirow{2}{*}{Kruskal–Wallis} & 0.741 & 0.984 & \textbf{1.000}  & \textbf{1.000}  & 0.938 & 0.999 & \textbf{1.000}  & \textbf{1.000}  \\ 
			[-2ex]& & (0.0579) & (0.0133) & (0.0000) & (0.0000) & (0.0240) & (0.0037)& (0.0000) & (0.0000)\\ 
			\cline{2-10} 
			& \multirow{2}{*}{ \texttt{DESeq2}} & 0.783 & \textbf{0.999} & \textbf{1.000}  & \textbf{1.000} & 0.731 & 0.916 & 0.980 & \textbf{1.000} \\ 
			[-2ex]& & (0.0461) & (0.0051) & (0.0000) & (0.0000) & (0.0606) & (0.0297) & (0.0177) & (0.0000)\\ 
			\cline{2-10} 
			& \multirow{2}{*}{ \texttt{edgeR}} & 0.717 & 0.987 & 0.999 & \textbf{1.000}  & 0.699 & 0.930 & 0.977 & \textbf{1.000} \\ 
			[-2ex]& & (0.0564) & (0.0127) & (0.0042) & (0.0000) & (0.0580) & (0.0392) & (0.0169) & (0.0000)\\ 
			\cline{2-10} 
			& \multirow{2}{*}{ \texttt{metagenomeSeq}} & 0.659 & 0.996 & 0.982 & 1.000 & \textbf{0.746} & \textbf{0.970} & 0.995 & \textbf{1.000} \\ 
			[-2ex]& & (0.0666) & (0.0082) & (0.0128) & (0.0000) & (0.0528) & (0.0238) & (0.0093) & (0.0000)\\ 
			\cline{1-10} 
			\multirow{13}{*}{ZINB} & \multirow{2}{*}{ZINB-DPP} & \textbf{0.459} & \textbf{0.845} & \textbf{0.912} & 0.971 & 0.716 & 0.705 & 0.956 & 0.986\\
			[-2ex]& & (0.0582) & (0.0467) & (0.0267) & (0.0186) & (0.0514) & (0.0532) & (0.0223) & (0.0142)\\ 
			\cline{2-10} 
			& \multirow{2}{*}{ZINB-TSS} & 0.403 & 0.835 & 0.906 & 0.969 & 0.681 & 0.704 & 0.954 & 0.986\\ 
			[-2ex]& & (0.0651) & (0.0402) & (0.0289) & (0.0167) & (0.0487) & (0.0520) & (0.0216) & (0.0144)\\ 
			\cline{2-10} 
			& \multirow{2}{*}{ZINB-Q75} & 0.407 & 0.837 & 0.906 & \textbf{0.972} & 0.676 & 0.704 & 0.952 & 0.985\\ 
			[-2ex]& & (0.0598) & (0.0428) & (0.0314) & (0.0187) & (0.0559) & (0.0520) & (0.0210) & (0.0157)\\ 
			\cline{2-10} 
			& \multirow{2}{*}{ZINB-RLE} & NA & NA &NA  & NA  & NA  & NA  & NA  & NA  \\ 
			[-2ex]& & (-) & (-) & (-) & (-) & (-) & (-) & (-) & (-)\\ 
			\cline{2-10} 
			& \multirow{2}{*}{ZINB-TMM} & 0.399 & 0.832 & 0.905 & 0.967 & 0.674 & 0.703 & 0.955 & \textbf{0.987} \\ 
			[-2ex]& &(0.0548) & (0.0448) & (0.0315) & (0.0185) & (0.0551) & (0.0520) & (0.0223) & (0.0138)\\ 
			\cline{2-10} 
			& \multirow{2}{*}{ZINB-CSS} &0.399 & 0.832 & 0.904 & 0.969 & 0.663 & 0.703 & 0.954 & 0.986 \\ 
			[-2ex]& & (0.0580) & (0.0433) & (0.0292) & (0.0190) & (0.0540) & (0.0522) & (0.0212) & (0.0125)\\ 
			\cline{2-10} 
			& \multirow{2}{*}{DM} &0.077 & 0.327 & 0.472 & 0.923 & 0.217 & 0.518 & 0.767 & 0.957 \\ 
			[-2ex]& &(0.0471) & (0.0553) & (0.0594) & (0.0277) & (0.0689) & (0.0593) & (0.0466) & (0.0243)\\ 
			\cline{1-10} 
		\end{tabular}
	}
\end{table}

\begin{table}
	\centering
	\resizebox{\columnwidth}{!}{%
		\begin{tabular}{c|c|c|c|c|c|c|c|c|c}
			\hline
			\hline
			\multirow{4}{*}{\pbox{20cm}{\shortstack{Generative \\ Model}}}
			& \multirow{4}{*}{Methods} & \multicolumn{8}{c}{Simulation Setting} \\
			\cline{3-10} 
			& & \multicolumn{4}{c|}{$K =2$} & \multicolumn{4}{c}{$K =3$} \\
			\cline{3-10} 
			& & \multicolumn{2}{c}{$n =24$} & \multicolumn{2}{c|}{$n =108$} & 
			\multicolumn{2}{c}{$n =24$} & \multicolumn{2}{c}{$n =108$}  \\
			\cline{1-10} 
			& & $\sigma = 1$ & $\sigma = 2$ & $\sigma = 1$ & $\sigma = 2$ & $\sigma = 1$ & $\sigma = 2$ & $\sigma = 1$ & $\sigma = 2$ \\
			\cline{3-10} 
			\multirow{6}{*}{ZINB} & \multirow{2}{*}{ANOVA} & 0.099 & 0.412 & 0.741 & 0.930 & 0.217 & 0.374 & 0.870 & 0.954 \\ 
			[-2ex]& & (0.0720) & (0.0847) & (0.0511) & (0.0234) & (0.1070) & (0.0917) & (0.0341) & (0.0271)\\ 
			\cline{2-10} 
			& \multirow{2}{*}{Kruskal–Wallis} & 0.031 & 0.122 & 0.372 & 0.634 & 0.007 & 0.010 & 0.488 & 0.795\\ 
			[-2ex]& & (0.0583) & (0.0862) & (0.0689) & (0.0557) & (0.0347) & (0.0392) & (0.0726) & (0.0435)\\ 
			\cline{2-10} 
			& \multirow{2}{*}{ \texttt{DESeq2}} & 0.223 & 0.627 & 0.727 & 0.963 & 0.295 & 0.408 & 0.643 & 0.844\\ 
			[-2ex]& & (0.0580) & (0.0672) & (0.0554) & (0.0237) & (0.0685) & (0.0647) & (0.0677) & (0.0420)\\ 
			\cline{2-10} 
			& \multirow{2}{*}{ \texttt{edgeR}} & NA & NA &NA  & NA  & NA  & NA  & NA  & NA  \\ 
			[-2ex]& & (-) & (-) & (-) & (-) & (-) & (-) & (-) & (-)\\ 
			\cline{2-10} 
			& \multirow{2}{*}{ \texttt{metagenomeSeq}} & 0.072 & 0.232 & 0.223 & 0.672 & 0.094 & 0.154 & 0.186 & 0.360\\ 
			[-2ex]& & (0.0518) & (0.0598) & (0.0603) & (0.0675) & (0.0506) & (0.0743) & (0.0560) & (0.0625)\\ 
			\hline
			\hline
		\end{tabular}
	}
	\caption{DM and ZINB simulation: Matthews correlation coefficient (MCC) given by all methods. In each cell, the top number is the averaged MCC over 50 independent datasets, and the bottom number in parentheses is the standard error. The result from the model that achieved best performance under the associated scenario (each column) is marked in bold.}
	\label{simumcc},
\end{table}

\clearpage
\begin{table}
	\centering
	\resizebox{\columnwidth}{!}{%
		\begin{tabular}{c|c|c|c|c|c|c|c|c|c}
			\hline
			\hline
			\multirow{4}{*}{\pbox{20cm}{\shortstack{Real Data\\ Sample Type}}}
			& \multirow{4}{*}{Methods} & \multicolumn{8}{c}{Synthetic Setting} \\
			\cline{3-10} 
			& & \multicolumn{4}{c|}{AUC} & \multicolumn{4}{c}{MCC} \\
			\cline{3-10} 
			& & \multicolumn{2}{c}{$\log(\sigma)=1$} & \multicolumn{2}{c|}{$\log(\sigma)=2$} & 
			\multicolumn{2}{c}{$\log(\sigma)=1$} & \multicolumn{2}{c}{$\log(\sigma)=2$}  \\
			\cline{1-10} 
			& & $n=24$ & $n=108$ & $n=24$ & $n=108$ & $n=24$ & $n=108$ & $n=24$ & $n=108$ \\
			\cline{3-10} 
			\multirow{13}{*}{Skin} & \multirow{2}{*}{ZINB-DPP} & 0.925 & 0.965 & \textbf{0.994} & 0.998 & \textbf{0.662} &\textbf{0.830} & \textbf{0.928} & \textbf{0.989} \\
			[-2ex]& & (0.0310) & (0.0199) & (0.0074) & (0.0047) & (0.0736) & (0.0613) & (0.0469) & (0.0155)\\ 
			\cline{2-10} 
			& \multirow{2}{*}{ZINB-TSS} & 0.923 & 0.957 & 0.991 & 0.999 & 0.670 & 0.825 & 0.920 & 0.983  \\ 
			[-2ex]& & (0.0293) & (0.0259) & (0.0110) & (0.0041) & (0.0727) & (0.0591) & (0.0469) & (0.0214)\\ 
			\cline{2-10} 
			& \multirow{2}{*}{ZINB-Q75} & 0.920 & 0.959 & 0.991 & 0.998 & 0.658 & 0.813 & 0.920 & 0.986\\ 
			[-2ex]& & (0.0335) & (0.0247) & (0.0109) & (0.0036) & (0.0704) & (0.0756) & (0.0450) & (0.0190)\\ 
			\cline{2-10} 
			& \multirow{2}{*}{ZINB-RLE} & 0.923 & 0.952 & 0.990 & 0.998 & 0.658 & 0.825 & 0.921 & 0.982 \\ 
			[-2ex]& &(0.0249) & (0.0277) & (0.0117) & (0.0042) & (0.0723) & (0.0609) & (0.0436) & (0.0215)\\ 
			\cline{2-10} 
			& \multirow{2}{*}{ZINB-TMM} & 0.925 & 0.952 & 0.990 & 0.998 & 0.658 & 0.825 & 0.921 & 0.986 \\ 
			[-2ex]& & (0.0271) & (0.0274) & (0.0099) & (0.0053) & (0.0786) & (0.0608) & (0.0467) & (0.0203) \\ 
			\cline{2-10} 
			& \multirow{2}{*}{ZINB-CSS} & 0.909 & 0.956 & 0.988 & 0.998 & 0.640 & 0.822 & 0.914 & 0.986 \\ 
			[-2ex]& & (0.0348) & (0.0233) & (0.0116) & (0.0047) & (0.0857) & (0.0752) & (0.0491) & (0.0220)\\ 
			\cline{2-10} 
			& \multirow{2}{*}{DM} & \textbf{0.929} & \textbf{0.978} & \textbf{0.994} & 1.000 & 0.639 & 0.819 & \textbf{0.928} & 0.985 \\ 
			[-2ex]& &( 0.0246) & (0.0124) & (0.0060) & (0.0011) & (0.0684) & (0.0480) & (0.0418) & (0.0172) \\ 
			\cline{2-10} 
			& \multirow{2}{*}{ANOVA} & 0.851 & 0.946 & 0.976 & 0.998 & 0.579 & 0.744 & 0.831 & 0.960 \\ 
			[-2ex]& &(0.0528) & (0.0263) & (0.0179) & (0.0046) & (0.0884) & (0.0753) & (0.0635) & (0.0277)\\ 
			\cline{2-10} 
			& \multirow{2}{*}{Kruskal–Wallis} & 0.846 & 0.966 & 0.979 & \textbf{1.000} & 0.572 & 0.787 & 0.844 & 0.983\\ 
			[-2ex]& & (0.0557) & (0.0215) & (0.0154) & (0.0005) & (0.0968) & (0.0692) & (0.0635) & (0.0182)\\ 
			\cline{2-10} 
			& \multirow{2}{*}{ \texttt{DESeq2}} & 0.770 & 0.866 & 0.951 & 0.990 & 0.640 & 0.690 & 0.890 & 0.958\\ 
			[-2ex]& & (0.0570) & (0.0443) & (0.0327) & (0.0138) & (0.0569) & (0.0878) & (0.0630) & (0.0349)\\ 
			\cline{2-10} 
			& \multirow{2}{*}{ \texttt{edgeR}} & 0.768 & 0.919 & 0.957 & 0.996 & 0.481 & 0.668 & 0.795 & 0.957\\ 
			[-2ex]& & (0.0694) & (0.0382) & (0.0316) & (0.0073) & (0.1498) &( 0.1098) & (0.1004) & (0.0342)\\ 
			\cline{2-10} 
			& \multirow{2}{*}{ \texttt{metagenomeSeq}} & 0.637 & 0.953 & 0.971 & 1.000 & 0.558 & 0.704 & 0.813 & 0.936\\ 
			[-2ex]& &(0.0783) & (0.0242) & (0.0195) & (0.0006) & (0.0863) & (0.0592) & (0.0703) & (0.0346)\\ 
			\cline{1-10} 
			\multirow{13}{*}{Feces} & \multirow{2}{*}{ZINB-DPP} & \textbf{0.917} & 0.891 & \textbf{0.987} & 0.979 & 0.619 & \textbf{0.658} & \textbf{0.884} & 0.900 \\
			[-2ex]& & (0.0294) & (0.0371) & (0.0162) & (0.0193) & (0.0974) & (0.0837) & (0.0831) & (0.0518)\\ 
			\cline{2-10} 
			& \multirow{2}{*}{ZINB-TSS} &0.900 & 0.857 & 0.975 & 0.968 & 0.620 & 0.627 & 0.872 & 0.883\\ 
			[-2ex]& & (0.0370) & (0.0499) & (0.0222) & (0.0237) & (0.1003) & (0.0977) & (0.0810) & (0.0552)\\ 
			\cline{2-10} 
			& \multirow{2}{*}{ZINB-Q75} & 0.874 & 0.858 & 0.970 & 0.966 & 0.553 & 0.625 & 0.861 & 0.879\\ 
			[-2ex]& & (0.0691) & (0.0482) & (0.0254) & (0.0247) & (0.1310) & (0.1009) & (0.0818) & (0.0634)\\ 
			\cline{2-10} 
			& \multirow{2}{*}{ZINB-RLE} & 0.909 & 0.863 & 0.975 & 0.962 & \textbf{0.630} & 0.630 & 0.868 & 0.876  \\ 
			[-2ex]& & (0.0346) & (0.0462) & (0.0235) & (0.0238) & (0.1028) & (0.0848) & (0.0826) & (0.0650)\\ 
			\cline{2-10} 
			& \multirow{2}{*}{ZINB-TMM} & 0.912 & 0.869 & 0.976 & 0.966 & 0.623 & 0.638 & 0.873 & 0.878 \\ 
			[-2ex]& &(0.0333) & (0.0435) & (0.0218) & (0.0235) & (0.0981) & (0.0921) & (0.0895) & (0.0574)\\ 
			\cline{2-10} 
			& \multirow{2}{*}{ZINB-CSS} &0.887 & 0.858 & 0.978 & 0.968 & 0.593 & 0.639 & 0.870 & 0.887 \\ 
			[-2ex]& & (0.0493) & (0.0564) & (0.0202) & (0.0262) & (0.1096) & (0.0904) & (0.0657) & (0.0594)\\ 
			\cline{2-10} 
			& \multirow{2}{*}{DM} & \textbf{0.917} & \textbf{0.929} & \textbf{0.987} & \textbf{0.993} & 0.594 & 0.655 & \textbf{0.884} & \textbf{0.928}\\ 
			[-2ex]& &(0.0295) & (0.0293) & (0.0137) & (0.0080) & (0.0982) & (0.0811) & (0.0764) & (0.0390)\\ 
			\cline{1-10} 
		\end{tabular}
	}
\end{table}

\begin{figure}[!h]
	\centering
	\includegraphics[width=.7\textwidth]{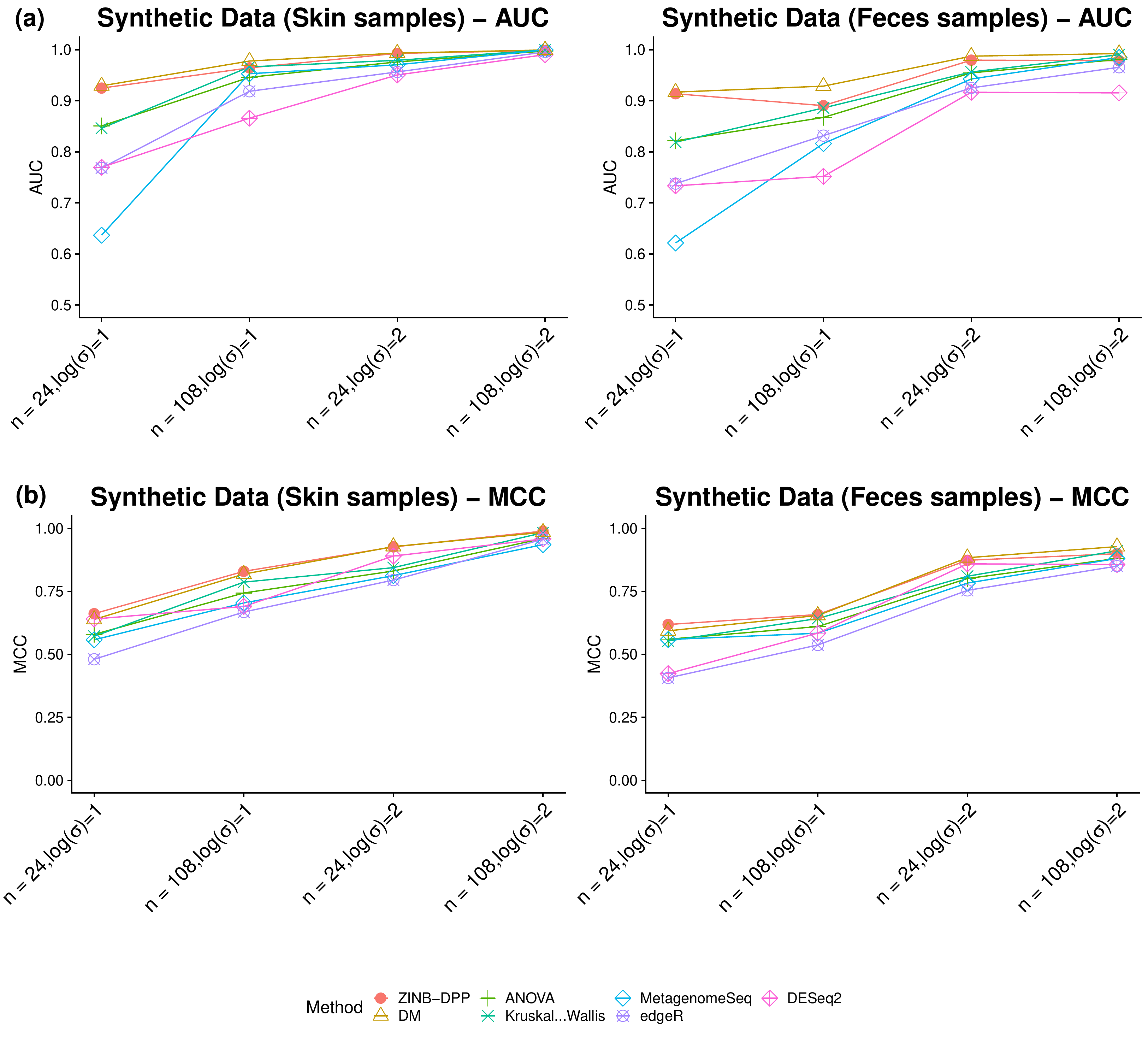} 
	\caption{Simulation study: The average AUC (a) and MCC (b) achieved by the proposed framework with DM model and ZINB model, and the five competitors: ANOVA, Kruskal-Wallis, \texttt{edgeR}, \texttt{DESeq2}, and \texttt{metagenomeSeq}. Results are plotted from the synthetic data generated by the multinational model of skin/feces samples.}
	\label{simu_fig_2}
\end{figure}

The ZINB-DPP model also shows very competitive performance on the synthetic data. The results of the average AUC and MCC are presented in Figure \ref{simu_fig_2}(a) and \ref{simu_fig_2}(b). Note that we generated the synthetic datasets from the multinomial model whose parameters were estimated by using the skin/feces samples collected by \cite{caporaso2011global}. Therefore, they ought to favor our DM model. However, the ZINB-DPP model, again, maintained the highest MCCs across all scenarios, and the DM model performed the second-best in general. Additionally, all methods showed great improvement when either the sample size or the effect size was increased, which was expected. The related numerical results are summarized in Table \ref{synth}, which compares the AUCs and MCCs for all methods implemented on the synthetic data.%We show only the results from skin samples here; and the plots of feces, which lead to the same findings, are shown in the supplement (where ?? XX). Results from the ZINB model with alternative normalization methods are also included in the supplement (where ?? XX). 

\begin{table}
	\centering
	\resizebox{\columnwidth}{!}{%
		\begin{tabular}{c|c|c|c|c|c|c|c|c|c}
			\hline
			\hline
			\multirow{4}{*}{\pbox{20cm}{\shortstack{Real Data\\ Sample Type}}}
			& \multirow{4}{*}{Methods} & \multicolumn{8}{c}{Synthetic Setting} \\
			\cline{3-10} 
			& & \multicolumn{4}{c|}{AUC} & \multicolumn{4}{c}{MCC} \\
			\cline{3-10} 
			& & \multicolumn{2}{c}{$\log(\sigma)=1$} & \multicolumn{2}{c|}{$\log(\sigma)=2$} & 
			\multicolumn{2}{c}{$\log(\sigma)=1$} & \multicolumn{2}{c}{$\log(\sigma)=2$}  \\
			\cline{1-10} 
			& & $n=24$ & $n=108$ & $n=24$ & $n=108$ & $n=24$ & $n=108$ & $n=24$ & $n=108$ \\
			\cline{3-10} 
			\multirow{7}{*}{Feces} & \multirow{2}{*}{ANOVA} & 0.822 & 0.867 & 0.955 & 0.981 & 0.560 & 0.610 & 0.801 & 0.881 \\ 
			[-2ex]& &(0.0636) & (0.0491) & (0.0339) & (0.0177) & (0.1116) & (0.0723) & (0.0865) & (0.0542)\\ 
			\cline{2-10} 
			& \multirow{2}{*}{Kruskal–Wallis} & 0.819 & 0.886 & 0.957 & 0.990 & 0.553 & 0.642 & 0.810 & 0.911\\ 
			[-2ex]& & (0.0611) & (0.0436) & (0.0354) & (0.0112) & (0.1077) & (0.0860) & (0.0980) & (0.0508) \\ 
			\cline{2-10} 
			& \multirow{2}{*}{ \texttt{DESeq2}} & 0.733 & 0.752 & 0.917 & 0.916 & 0.424 & 0.584 & 0.859 & 0.857\\ 
			[-2ex]& & (0.0860) & (0.0562) & (0.0590) & (0.0428) & (0.1573) & (0.1043) & (0.0901) & (0.0818)\\ 
			\cline{2-10} 
			& \multirow{2}{*}{ \texttt{edgeR}} & 0.738 & 0.832 & 0.925 & 0.966 & 0.407 & 0.537 & 0.754 & 0.851\\ 
			[-2ex]& & (0.0796) & (0.0556) & (0.0523) & (0.0270) & (0.1474) & (0.1035) & (0.1275) & (0.0705)\\ 
			\cline{2-10} 
			& \multirow{2}{*}{ \texttt{metagenomeSeq}} & 0.621 & 0.816 & 0.943 & 0.985 & 0.559 & 0.584 & 0.783 & 0.881\\ 
			[-2ex]& & (0.0892) & (0.0623) & (0.0426) & (0.0168) & (0.1013) & (0.0937) & (0.0860) & (0.0557)\\ 
			\hline
			\hline
		\end{tabular}
	}
	\caption{Synthetic data: area under the curve (AUC) and Matthews correlation coefficient (MCC) given by all methods. In each cell, the top number is the averaged AUC (or MCC) over 50 independent datasets, and the bottom number in parentheses is the standard error. The result from the model that achieved best performance under the associated scenario (each column) is marked in bold.}
	\label{synth}
\end{table}

\subsection*{S3.5 \quad  Sensitivity analysis}\label{sa}
We examined the model sensitivity with respect to the choice of hyperparameters $b_0, \ldots, b_K$ and $h_0, \ldots, h_K$. The results in Table \ref{sensi_table} show that our approach is considerably insensitive to the hyperparameter settings. 

The choice of $b_k$ and $h_k$ for $k = 0,\ldots, K$ are related to the variance terms in the Gaussian mixture model of the top-level. Large values of $h_k$ would achieve a noninformative prior on $\mu_{kj}$'s. On the other hand, as we specified IG($a_k,b_k$) prior for $\sigma_{kj}^2$ and set $a_k = 2$ for all $k = 0,\ldots, K$, the resulting variance of inverse gamma distribution does not exist. We considered a range of $(b_k, h_k)$ settings as $b_k \in \{0.1, 1, 2, 10\}$ and $h_k \in \{1, 10, 100\}$. Then we applied the ZINB-DPP model with different combinations of $(b_k, h_k)$ to datasets simulated from the ZINB model discussed in Section 4 in the main text. To fully assess the impact of hyperparameters under different scenarios, we considered $K = 2, 3$ and $n = 24, 108$ with a weakly discriminating signal $\sigma = 1$. We generated 50 independent datasets for each case and reported the averaged AUC (in Table \ref{sensi_table}). Clearly, the AUC remained stable for different choices of $(b_k, h_k)$. We suggest to set $b_k = 1$ and $h_k$ to be any value ranging from 10 to 100 for $k= 1, \ldots K$.
\begin{sidewaystable}
	%\begin{table}[h]
	\centering
	\small
	\caption{Sensitivity Analysis: AUC and the corresponding standard error (in parenthesis) for different choice of hyperparameters}\label{sensi_table}
	\begin{tabular}{c|c|c|c|c|c|c|c|c|c|c|c|c}
		\hline
		\hline
		\pbox{20cm}{\shortstack{$b_k$}} & \multicolumn{3}{c|}{0.1} & \multicolumn{3}{c|}{1} 
		& \multicolumn{3}{c|}{2} & \multicolumn{3}{c}{10} \\
		\cline{1-13}
		\pbox{20cm}{\shortstack{$h_k$}} & 1 & 10 & 100&  1 & 10& 100
		& 1 & 10& 100 & 1 & 10 & 100 \\
		\cline{1-13}
		$K = 2$ & 0.887 & 0.888 & 0.879 & 0.888 & 0.878 & 0.869 & 0.871 & 0.858 & 0.846 & 0.739 & 0.730 & 0.711\\
		[-2ex] $n=24$ & (0.0309) & (0.0253) & (0.0239) & (0.0244) & (0.0258) & (0.0243) & (0.0261) & (0.0290 )& (0.0268) & (0.0354) & (0.0336) & (0.0421)\\
		\cline{2-13}\\
		\cline{2-13}
		$K = 2$ & 0.987 & 0.986 & 0.981 & 0.996 & 0.994 & 0.993 & 0.996 & 0.996 & 0.995 & 0.967 & 0.964 & 0.946 \\
		[-2ex] $n=108$ & (0.0127)& (0.0136) & (0.0184) & (0.0057) & (0.0064) & (0.0069) & (0.0046) & (0.0051) & (0.0052) & (0.0098) & (0.0148) & (0.0193)\\
		\cline{2-13}\\
		\cline{2-13}
		$K = 3$ & 0.780 & 0.785 & 0.783 & 0.787 & 0.784 & 0.792 & 0.785 & 0.783 & 0.793 & 0.722 & 0.706 & 0.675 \\
		[-2ex] $n=24$ & (0.0562) & (0.0577) & (0.0589) & (0.0581) & (0.0569) & (0.0542) & (0.0576) & (0.0563) & (0.0503) & (0.0549) & (0.0511) & (0.0438)\\
		\cline{2-13}\\
		\cline{2-13}
		$K = 3$ & 0.992 & 0.991 & 0.983 & 0.998 & 0.997 & 0.996 & 0.998 & 0.998 & 0.997 & 0.994 & 0.986 & 0.936\\
		[-2ex] $n=108$ & (0.0104)& (0.0110) & (0.0163) & (0.0062) & (0.0067) & (0.0063) & (0.0053) & (0.0047) & (0.0057) & (0.0058) & (0.0103) & (0.0279)\\
		\hline 
		\hline
	\end{tabular}
\end{sidewaystable}

\section*{S4 \quad  Real Data Analysis}

\subsection*{S4.1 \quad   Quality control}

Before analyzing a given microbiome count dataset, we first implement a simple quality control step. It ensures that the dataset is of the best quality to perform the subsequent modeling. This step includes: 1) examining the total number of reads sequenced, and 2) verifying the richness of taxa discovered. In all, quality control is considered for both sample (patient) and feature (taxon) levels.

\subsubsection*{S4.1.1 \quad   Sample-wise quality control}

% todo %
%collector's curve citation%
In sequencing data analysis, if the total number of reads for a sample falls above or below specific values (shown below), then this may indicate poor sequence quality owing to duplicate reads or limited sampling bias. Specifically, let $y_i=\sum_{j=1}^py_{ij}$ denote the total number of reads observed in sample $i$. A sample $i$ will be removed if its total reads $y_i<\text{Q1}-3(\text{Q3}-\text{Q1})$ or $>\text{Q3}+3(\text{Q3}-\text{Q1})$, where $\text{Q1}$ and $\text{Q3}$ are the lower and upper quartiles (i.e. the $25$th and $75$th percentiles) of the total reads of all the samples (i.e. $\{y_1,\ldots,y_n\}$). Note that in the context of box-and-whisker plotting, a data point is defined as an extreme outlier if it stands outside these two limits. Second, in ecology, investigators find that the number of species increases as sampling effort increases. This species-abundance distribution can be depicted by \textit{the collector's curve}, which is monotonically increasing and negatively accelerated. Hence, we assumed that the logarithmic count of taxa discovered in one sample had a linear relationship with the total reads observed in the same sample. As suggested by \cite{Hair2006}, we fitted the regression model to compute the Cook's distance for each patient, and removed the ones with distances above $4/(n-2)$ since they were considered to be the influential data points for a least-squares regression analysis. 

\subsubsection*{S4.1.2 \quad   Feature-wise quality control}

Another common procedure in microbiome studies is to filter out the extremely low-abundance taxa. For example, \cite{wadsworth2017integrative} requires each genus in their model to be present in at least 5\% of the samples. Similarly, \cite{qin2014alterations} kept the taxa with median compositional abundance greater than 0.01\% of total abundance in either the healthy control group or the disease group. In our ZINB model, the estimation of the dispersion parameter of each feature (taxon) involves the calculation of the second moment, similar to computing the variance component in the Gaussian mixture model. Therefore, it requires at least two observed reads in each patient group to perform the analysis. In practice, we suggest removing a taxon if it has fewer than three nonzero reads in any patient group. In our second case study, we relax the threshold such that we removed taxa with fewer than two nonzero reads in any patient group, due to the small sample size ($n = 27$).

\subsection*{S4.2\quad   Comparison with alternative approaches}
Along with the simulation study conducted in the paper, we compared the results given by our proposed models (DM, ZINB-DPP) on the case study data with those from alternative approaches, including ANOVA, Kruskal–Wallis test, \texttt{DESeq2}, \texttt{edgeR} and \texttt{metagenomeSeq}.

\subsubsection*{S4.2.1 \quad   Colorectal cancer study}
\begin{sidewaysfigure}
	\centering
	\includegraphics[width = 0.8\linewidth]{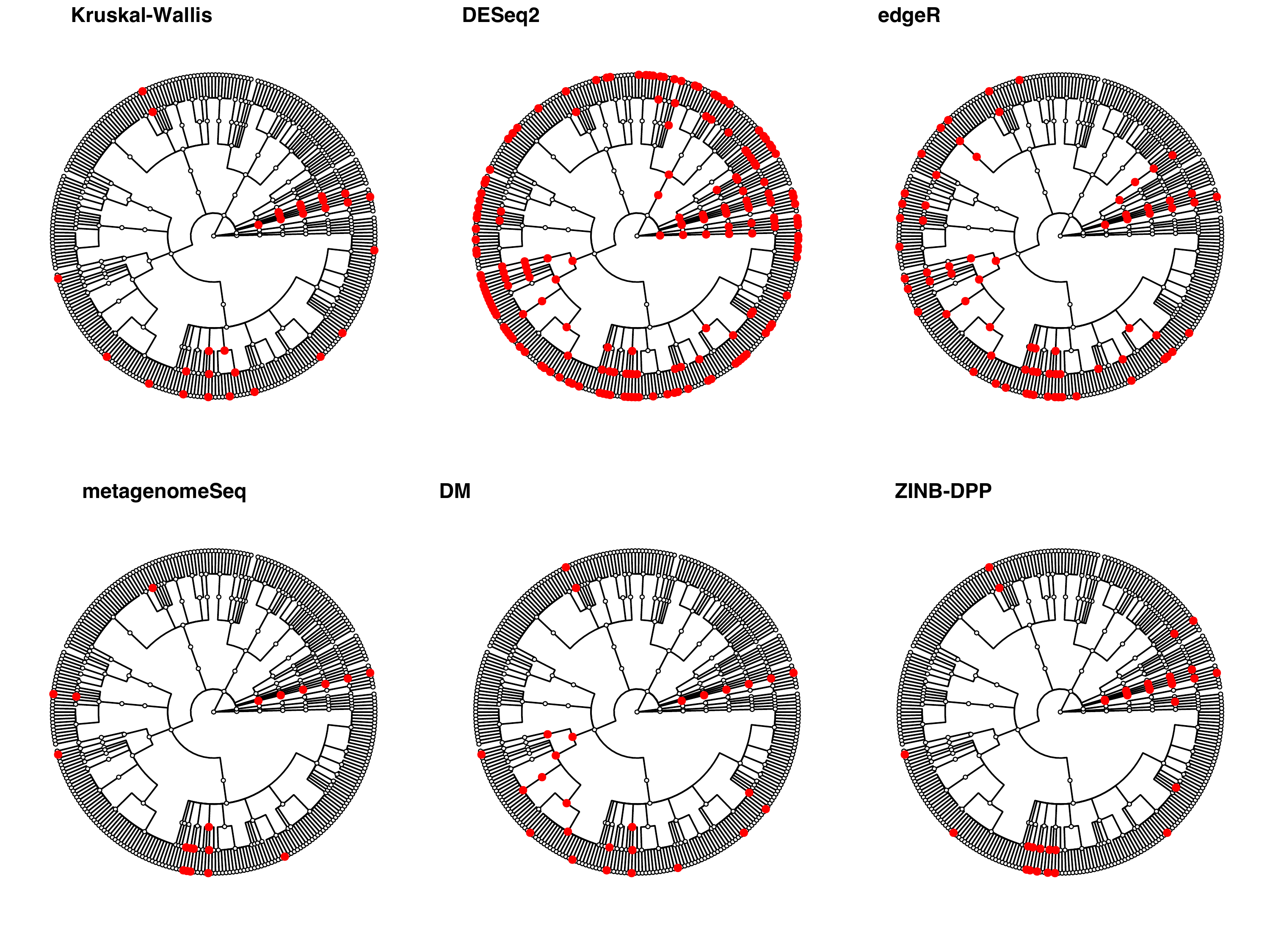}
	\cprotect\caption{Colorectal cancer study: the discriminating taxa identified by different methods. The red dots in each of the first four cases represent the taxa with Benjamini-Hochberg adjusted $p$-values below the significance level of $1\%$. The red dots in the DM and ZINB-DPP are taxa detected by controlling the Bayesian FDR to be less than $1\%$.}
	\label{tree_other_1}
\end{sidewaysfigure}

We adopted a 1\% significance level threshold on the Benjamini-Hochberg (BH) adjusted p-values provided by the alternative methods. The choice of 1\% was set to be consistent with the Bayesian false discovery rate (FDR) of the ZINB-DPP model. For the DM model, we kept the same hyperprior settings as for the ZINB-DPP model, i.e., we set $a_0=a_1=\ldots=a_k = 2$, $b_0=b_1=\ldots=b_k=1$ for variance components $\sigma_{0j}^2$ and $\sigma_{kj}^2$, and we let $h_0 = h_1 = \ldots = h_K = 50$. We further adopted the same Markov random field settings as $d = -2.2$ and $f = 0.5$. The results for the DM model were obtained by controlling the Bayesian FDR to be less than 1\%.  Figure \ref{tree_other_1} compares all the results. First, ANOVA lacked statistical power when the data contained too many zeros, and it failed to identify any discriminating taxa in this case. Therefore, Figure \ref{tree_other_1} excludes the result by ANOVA. The Kruskal–Wallis test identified 30 discriminating taxa, 19 of which were also reported by the ZINB-DPP model. Although Kruskal–Wallis selected the branch of species \textit{Fusobacterium nucleatum} as all the other methods did, it failed to detect the phylogenetic branch from \textit{Synergistaceae} to \textit{Synergistetes}, which was reported by the ZINB-DPP model and \textit{Synergistaceae} was found to be CRC-enriched in a previous study \citep{coker2019enteric}. Next, under a stringent significance level of 1\%, \texttt{DESeq2} and \texttt{edgeR} still led the selection of 179 and 72 discriminating taxa, respectively. The large number of detections might suggest a high FDR. Furthermore, \texttt{edgeR} failed to detect the phylogenetic branch from \textit{Synergistaceae} to its phylum level. Lastly, we found that \texttt{metagenomeSeq} and the DM model performed conservatively, as they only reported 20 and 27 discriminating taxa, respectively. 14 out of 20 taxa detected by \texttt{metagenomeSeq} were consistent with the result by ZINB-DPP, while 15 out of 27 findings from the DM model overlapped with results by the ZINB-DPP model. Although both of these methods reported \textit{Fusobacterium nucleatum} to be differentially abundant between two groups, neither of them detected the co-occurrence between \textit{Fusobacterium nucleatum} and \textit{Campylobacter}. 

Next, we focused on the species level detections by the ZINB-DPP model and the Kruskal–Wallis (KW) test. Under the Bayesian FDR or the significance level of $1\%$, the ZINB-DPP and the KW test reported 10 and 12 species, respectively, with seven species in common and 16 in total. For each of 16 species detected, as listed in Table 3 and 4 of the main text, we provided either the posterior probability of inclusion (PPI) or the BH adjusted p-value. The underlined PPI or p-value means that the species was selected as differentially abundant between two groups by the corresponding method. We conducted a comprehensive literature search to find biological evidence for each species listed in Table 3 and 4 of the main text. Six out of 11 species selected by our ZINB-DPP model were supported by previous studies, while there was no convincing evidence for the additional five species given by the KW test.

\subsubsection*{S4.1.2 \quad   Schizophrenia study}
\begin{sidewaysfigure}
	\centering
	\includegraphics[width = 0.8\linewidth]{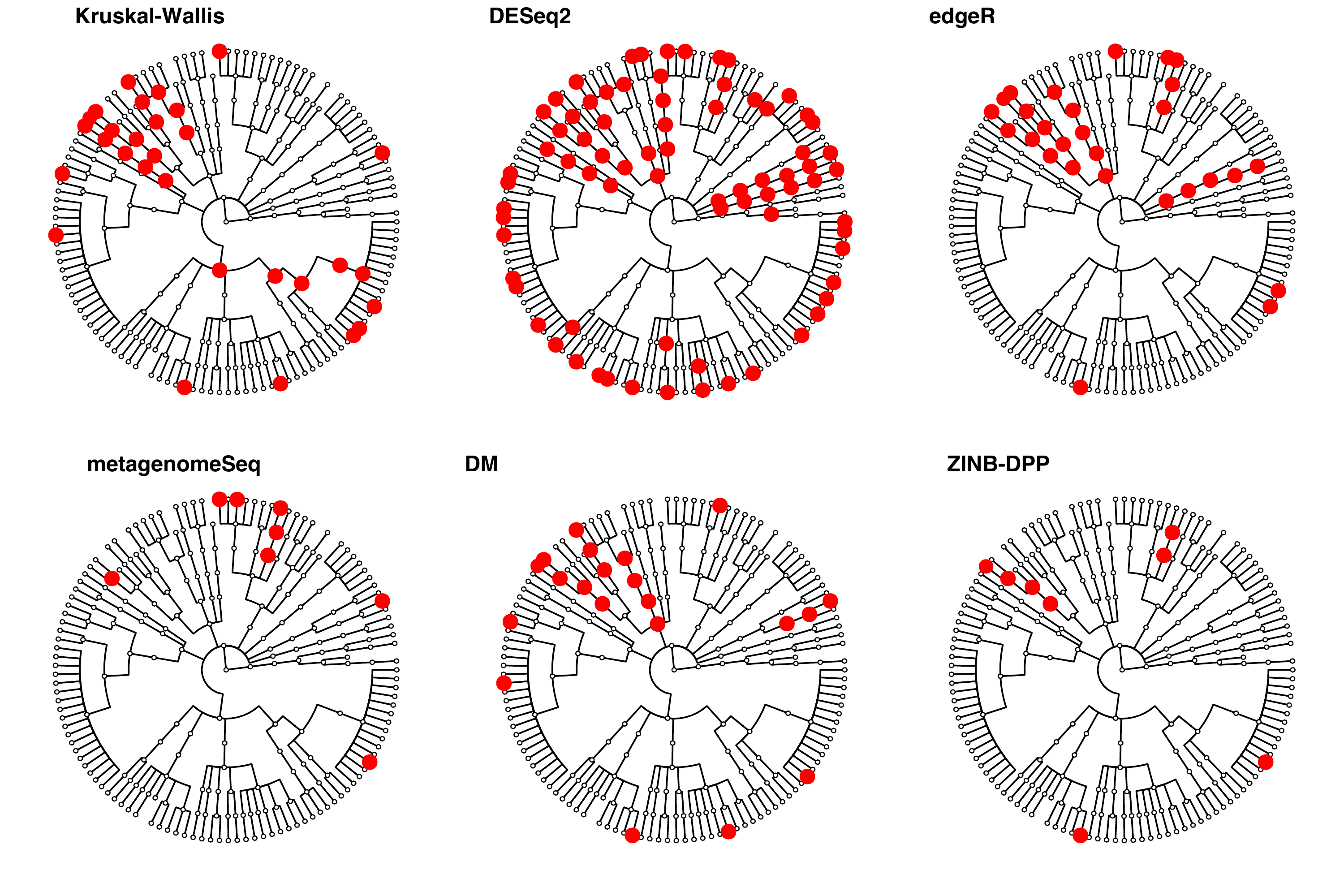}
	\cprotect\caption{Schizophrenia study : the discriminating taxa identified by different methods. The red dots in each of the first four cases represent the taxa with Benjamini-Hochberg adjusted $p$-values below the significance level of $5\%$. The red dots in the DM and ZINB-DPP are taxa detected by controlling the Bayesian FDR to be less than $5\%$.}
	\label{tree_other_2}
\end{sidewaysfigure}

Under a 5\% significance level threshold on the Benjamini-Hochberg adjusted p-values, we evaluated the performance of the alternative methods on the schizophrenia study. As for the DM model, we kept the same hyperprior settings as the ZINB model as described in the main text. The results are shown in Figure \ref{tree_other_2}. All the methods were challenged by the the small sample size ($n = 27$), along with the inflated amount of zeros. Kruskal–Wallis test, \texttt{DESeq2}, \texttt{edgeR} and the DM model leaded to the selection of 30, 81, 29 and 31 discriminating taxa, while the ANOVA test again failed to report any results. Out of the taxa selected by the Kruskal–Wallis test, \texttt{DESeq2}, \texttt{edgeR} and the DM model, respectively, five, eight, seven and five were in the list of taxa found by our model (ZINB-DPP selected eight under the Bayesian FDR of 5\%). One out of the eight taxa identified by our method but not \texttt{edgeR} was also selected by \texttt{DESeq2} and \texttt{metagenomeSeq}. Kruskal–Wallis test and the DM model, though they already reported about 30 differentially abundance taxa, failed to include the phylogenetic tree branch from \textit{Corynebacterium} to \textit{Corynebacteriaceae}, which were detected by \texttt{DESeq2}, \texttt{metagenomeSeq} and ZINP-DPP model. \texttt{metagenomeSeq} identified 9 taxa under the significance level of 5\%, five of which were consistent with the ZINB-DPP model. We noticed that \texttt{metagenomeSeq} only identified \textit{Neisseria sp.} and \textit{Neisseria} as in the phylogenetic tree branch, whereas all the remaining methods reached to the order level \textit{Neisseriales}. Meanwhile, \textit{Veillonella parvula} reported by ZINB-DPP, \texttt{DESeq2} and \texttt{edgeR} was not identified by \texttt{metagenomeSeq}.

%\section{BibTeX}
\bibliographystyle{agsm}
{\footnotesize
	\bibliography{JASA_manus_0413.bib}}
%\bibliography{Bibliography-MM-MC}

\end{document}